\documentclass{iopart}
\usepackage{epsfig}
\usepackage{amssymb}

\def\MET{\mbox{${\hbox{$E$\kern-0.6em\lower-.1ex\hbox{/}}}_T$}} 

\def\MP{\mbox{$M_D$}}
\def\mp{\mbox{$M_D$}\ }
\def\TH{\mbox{$T_H$}\ }
\def\mbh{\mbox{$M_{\rm BH}$}\ }     
\def\MBH{\mbox{$M_{\rm BH}$}}         
\def\MET{\mbox{${\hbox{$E$\kern-0.6em\lower-.1ex\hbox{/}}}_T$}} 
\def\met{\mbox{${\hbox{$E$\kern-0.6em\lower-.1ex\hbox{/}}}_T$}\ } 
\def\ifb{fb$^{-1}$}                     

\def\kMPl{\mbox{$k/\overline M_{\rm Pl}$}}

\def\D0{{D\O}\ }
\begin{document}

\topical{Black holes at future colliders and beyond}

\author{Greg Landsberg$^{\dagger}$}
\address{$^\dagger$Brown University, Department of Physics, 182 Hope St, Providence, RI 02912, USA.}
\ead{landsberg@hep.brown.edu}

\begin{abstract}
One of the most dramatic consequences of low-scale ($\sim 1~\mbox{TeV}$) quantum gravity in models with large or warped extra dimension(s) is copious production of mini black holes at future colliders and in ultra-high-energy cosmic ray collisions. Hawking radiation of these black holes is expected to be constrained mainly to our three-dimensional world and results in rich phenomenology. In this topical review we discuss the current status of astrophysical observations of black holes and selected aspects of mini black hole phenomenology, such as production at colliders and in cosmic rays, black hole decay properties, Hawking radiation as a sensitive probe of the dimensionality of extra space, as well as an exciting possibility of finding new physics in the decays of black holes.
\end{abstract}
\submitted{\JPG}
\maketitle

\smallskip

\section{Introduction}

In this topical review we briefly discuss the status of astronomical black hole observations and focus on phenomenology of the mini-black-hole production and decay in high-energy collisions in models with low-scale gravity. We point out exciting ways of studying quantum gravity and searching for new physics using large samples of black holes that may be accessible at future colliders and discuss the potential of the existing and future cosmic ray detectors for searches for black hole production in ultra-high-energy cosmic rays.

In what follows we will use ``natural'' units:
$$
	\hbar = c = k_B = 1,
$$
where $\hbar$ is the reduced Planck's constant, $c$ is the speed of light, and $k_B$ is Boltzmann's constant. This choice allows us to measure both energy and temperature in the same units, TeV ($= 10^{12}$~eV), while distance and time are measured in TeV$^{-1}$. The relationship between natural and SI units is as follows: 1~TeV$ = 1.16 \times 10^{16}$~K, 1~TeV$^{-1} = 6.58 \times 10^{-28}$~s~$= 1.97 \times 10^{-19}$~m.

\section{The hierarchy problem}

One of the most pressing problems in modern particle physics is the hierarchy problem, which can be expressed by a single, deceptively simple question: why is gravity (at least as we observe it) some $10^{38}$ orders of magnitude weaker than other forces of Nature? This mysterious fact is responsible for a humongous hierarchy of energy scales: from the scale of electroweak symmetry breaking ($M_{\rm EW} \sim 1$ TeV) to the energies at which gravity is expected to become as strong as other three forces (strong, electromagnetic, and weak), known as the Planck scale, or $M_{\rm Pl} = 1/\sqrt{G_N}\approx 1.22 \times 10^{16}$~TeV, where $G_N$ is Newton's coupling constant. 

If it was not for the hierarchy problem, our current understanding of fundamental particles and forces, consolidated in the standard model (SM) of particle physics, would have been nearly perfect. After all, the standard model has an impressive calculational power. Its predictions have been tested with a per mil or better accuracy, with hardly any significant deviations found so far. While certain phenomena, such as neutrino masses or existence of dark matter, are not explained in the standard model, they still may be accommodated with its minimum expansion, just like $CP$-violation has been earlier added to the SM framework via the quark mixing matrix. While it is true that several phenomena, including $CP$-violation or the hierarchy of fermion masses, are not really explained within the standard model, nevertheless they are readily accommodated within the SM by introducing additional parameters, e.g. fermion couplings to the Higgs field. Even the only remaining unobserved particle predicted in the standard model -- a fundamental Higgs boson responsible for the weak boson masses, while leaving photon massless -- may soon be found at the Large Hadron Collider (LHC, currently under construction at CERN, near Geneva, Switzerland) or even at the Tevatron (Fermilab, near Chicago, USA). Consequently, there are no fundamental experimental reasons to believe that the standard model cannot work to the highest energies, such as Planck scale. Nevertheless, there is a pretty good theoretical reason to believe that this is not the case: the hierarchy problem.

The reason that the hierarchy problem is so annoying stems from the fact that in our everyday experience large hierarchies tend to collapse, unless they are supported by some intermediate layers or by precisely tuned initial conditions (the so-called ``fine tuning''). For example, while it is mechanically possible to balance a pen on its point, the amount of fine tuning required for such a delicate balance is simply too demanding to be of practical use. Indeed, walking in a room and seeing a pen standing vertically on its point on a table would be too odd not to suspect some kind of a hidden support. Similarly, making the standard model work for all energies up to the Planck scale would require tremendous amount of fine tuning of its parameters~-- to a precision of $\sim (M_{\rm Pl}/M_{\rm EW})^2 \sim 10^{-32}$! A natural reaction to this observation is to conclude that the standard model needs some ``support,'' which should come either in a form of new physics at intermediate energy scales that adds extra layers to the hierarchy, or from certain new symmetries that would guarantee necessary amount of fine tuning.

However, it is worth pointing out that large amounts of fine tuning, while unnatural, are not prohibited by any fundamental principle. The fact that large hierarchies tend to collapse is merely empirical. While many large hierarchies we know have indeed collapsed, some other are surprisingly stable~-- examples of both kinds can be found in mechanical, social, and political systems alike. Certain examples of fine tuning have been observed in Nature: from the fact that the apparent angular size of the moon is the same as the angular size of the Sun within 2.5\%~-- a mere coincidence to which we owe such a spectacular sight as a solar eclipse~-- to a somewhat less
abused example of the infamous Florida 2000 presidential election recount with the ratio of Republican to Democratic votes equal to 1.000061, i.e. fine-tuned to unity with the precision of 0.006\%!

Despite these observations, the majority of physicists have been working very hard to find a more rational explanation of the hierarchy of forces in the standard model since its formulation in the late sixties. A number of viable solutions have been proposed as a result of this work: from supersymmetry (which protects the hierarchy due to nearly exact intrinsic cancelations of the effects caused by the standard model particles and by their superpartners obeying different spin statistics) to models with strong dynamics (which introduce an intermediate energy scale via new, QCD-like force, and often result in a composite Higgs boson).

Staying somewhat aside and quite debatable and controversial is an explanation of the standard model hierarchy via the anthropic principle: our universe is so fine tuned, because it happens to be one of a very few possible universes that could support intelligent life capable of raising this very question. Sparkled by the recent findings that the number of possible string theory vacua may be tremendously large: say, $10^{500}$ or may be even many orders of magnitude more, the anthropists claim that this vast landscape can give rise to enormous variety of possible universes, thus making it very likely the existence of significantly fine-tuned ones, which could support formation of heavy elements, stars, galaxies, and other precursors of intelligent life. For discussion of the recent controversy around the anthropic principle, see~\cite{anthropic}.

While each of the aforementioned solutions to the hierarchy problem easily deserves a separate review, here we will focus on yet another, more recently proposed remedy that does not involve new particles or symmetries, but instead uses the geometry of space itself.

\section{A brief history of space}

The idea that our space may contain more than the three familiar dimensions has been one of the most popular recurring themes in the work of numerous philosophers, artists, and writers since ancient times. From the very concept of Heaven and Hell to shadow or mirror worlds and parallel universes, the believe that the space around us may be a bigger place than it is commonly thought became one of the most mesmerizing and puzzling ideas, which inspired many great minds of the past centuries. However, it took quite a while before this concept has been truly embraced by scientists. 

In fact, in the beginning of the nineteenth century, following the seminal work by Gauss, there have been numerous publications that claimed that additional dimensions in space would contradict the inverse square law obeyed by gravitational and electromagnetic forces. Consequently the very concept of extra dimensions was quickly abandoned by mathematicians and physicists and remained the realm of art and philosophy.

Perhaps the first person to consider the possibility of extra dimensions in space rigorously was Bernhard Riemann~\cite{Riemann}. What started as an abstract mathematical idea of a curved Riemannian space, soon became the foundation of the most profound physics theory of the last century, if not of the entire history of physics: Albert Einstein's general relativity~\cite{GR}. While Einstein's theory was formulated in the three-plus-one space-time dimensions, it soon became apparent that the theory cannot be self-consistent up to the highest energies in its original form. 

In the 1920s, Theodor Kaluza and Oskar Klein~\cite{KK} suggested that a unification of electromagnetism and general relativity is possible if the fifth, spatial dimension of a finite size is added to the four-dimensional space-time. While this attempt has not led to a satisfactory and self consistent unification of gravity and electromagnetism, the idea of finite (or ``compactified'') extra spatial dimensions has been firmly established by Kaluza and Klein and eventually led to their broad use in string theory. Rapid progress in string theory in the 1970s helped the original idea of half-a-century earlier to regain its appeal. It was realized that extra six or seven spatial dimensions are required for the most economical and symmetric formulation of string theory. In particular, string theory requires extra dimensions to establish its deep connection with the supersymmetry, which also leads to the unification of gauge forces. While the size of these compact dimensions is not fixed in string theory, it is natural to expect them to have the radii similar to the inverse of the grand unification energy scale, or of the order of $10^{-32}$~m. Unfortunately, no experimental means exist to probe such short distances, so extra dimensions of string theory will likely remain untested even if this theory turns out to be the correct description of quantum gravity.

\section{Large extra spatial dimensions}

The situation with seemingly untestable extra spatial dimensions changed dramatically in the late 1990s, when Arkani-Hamed, Dimopoulos, and Dvali (ADD) suggested a new paradigm~\cite{add} in which several ($n$) of compactified extra dimensions could be as large as $\sim 1$~mm! These {\it large extra dimensions\/} have been introduced to solve the hierarchy problem of the standard model by dramatically lowering the Planck scale from its apparent value of $M_{\rm Pl} \sim 10^{16}$~TeV to $\sim 1$~TeV. (We further refer to this low {\it fundamental\/} Planck scale as \MP.) In the ADD model, the {\it apparent\/} Planck scale $M_{\rm Pl}$ only reflects the strength of gravity from the point of view of a three-dimensional (3D) observer. It is, in essence, just a virtual ``image'' of the fundamental, $(3+n)$-dimensional Planck scale $\MP$, caused by an incorrect interpolation of the gravitational coupling (measured only at low energies and large distances) to a completely different regime of high energies and short distances.

In order for such ``large'' extra dimensions not to violate any constraints from atomic physics and other experimental data, all other forces except for gravity must not be allowed to propagate in extra dimensions. Conveniently, modern quantum field theory allows to confine spin 0, 1/2, and 1 particles to a subset of a multidimensional space, or a ``brane.'' At the same time, the graviton, being a spin-2 particle, is not confined to the brane and thus permeates the entire (``bulk'') space. The solution to the hierarchy problem then becomes straightforward: if the Planck scale in the multidimensional space is of the order of the electroweak scale, there is no large hierarchy and no fine-tuning problem to deal with, as non-trivial physics really ``ends'' at the energies $\sim 1$~TeV. Gravity appears weak to a 3D-observer merely because it spans the entire bulk space, so the gravitons spend very little time in the vicinity of our brane. Consequently, the strength of gravity in 3D is literally diluted by an enormous volume of extra space.

The compactification radius of large extra dimensions ($R_c$) is fixed by their number and the value of the fundamental Planck scale in the ($4+n$)-dimensional space-time. By applying Gauss's law, one finds~\cite{add}
\begin{equation}
        M^2_{\rm Pl} = 8\pi M_D^{n+2}\, R_c^n.
\end{equation}
Here for simplicity we assumed that all $n$ extra dimensions have the same size $R_c$ and are compactified on a torus.

If one requires $M_D \sim 1$~TeV and a single extra dimension, its size has to be as large as the radius of the solar system; however, already for two extra dimensions their size is just $\sim 1$~mm; for $n=3$ it is $\sim 1$~nm, i.e. similar to the size of an atom; and for larger $n$ the radius further decreases to a subatomic size and reaches $\sim 1$~fm (the size of a proton) for seven extra dimensions.

A direct consequence of such compact extra dimensions in which gravity is allowed to propagate, is a modification of Newton's law at the distances comparable with $R_c$: the gravitational potential would fall off as $1/r^{n+1}$ for $r \lesssim R_c$. This immediately rules out the possibility of a single extra dimension, as the very existence of our solar system requires the potential to fall off as $1/r$ at the distances comparable with the size of planetary orbits. However, as of 1998 any higher number of large extra dimensions have not been ruled out by gravitational measurements, as Newton's law has not been tested to the distances smaller than about 1~mm. As amazing as it sounds, large extra dimensions with the size as macroscopic as $\sim 1$~mm were perfectly consistent with the host of experimental measurements at the time when the original idea has appeared!

Note that the solution to the hierarchy problem via large extra dimensions is not quite rigorous. In a sense, the hierarchy of energy scales is traded for the hierarchy of distance scales from the ``natural'' electroweak symmetry breaking range of 1 TeV$^{-1} \sim 10^{-19}$~m to the much larger size of compact extra dimensions. Nevertheless, this hierarchy could be significantly smaller than the hierarchy of energy scales or perhaps even be stabilized by some topological means. Strictly speaking, the idea of large extra dimensions is really a paradigm, which can be used to build more or less realistic models, rather than a model by itself.

Since large extra dimensions as a possible solution for the hierarchy problem were suggested, numerous attempts to either find them or to rule out the model have been carried out. They include measurements of gravity at sub-millimeter distances~\cite{tabletop}, studies of various astrophysical and cosmological implications~\cite{astro}, and numerous collider searches for virtual and real graviton effects~\cite{collider}. For a detailed review of the existing constraints and sensitivity of future experiments, the reader is referred to~\cite{collider,hs,gl}. The host of experimental measurements to date have largely disfavoured the case of two large extra dimensions. However for three or more extra dimensions, the lower limits on the fundamental Planck scale are only $\sim 1$~TeV, i.e. most of its allowed range has not been probed experimentally yet.

\section {The Randall-Sundrum model}

A significantly different and perhaps more rigorous solution to the hierarchy problem is offered in the Randall-Sundrum (RS) model~\cite{RS} with non-factorizable geometry and a single compact extra dimension. This is achieved by placing two 3-dimensional branes with equal and opposite tensions at the fixed points of the $S_1/Z_2$ orbifold in the five-dimensional anti-deSitter space-time ($AdS_5$). The metric of the $AdS_5$ space is given by $ds^2 = \exp(-2kR_c|\varphi|)\eta_{\mu\nu}dx^\mu dx^\nu - R_c^2 d\varphi^2$, where $0 \le |\varphi| \le \pi$ is the coordinate along the compact dimension of radius $R_c$, $k$ is the curvature of the $AdS_5$ space, often referred to as the warp factor, $x^\mu$ are the conventional (3+1)-space-time coordinates, and $\eta^{\mu\nu}$ is the metric tensor of Minkowski space-time. The positive-tension (Planck) brane is placed at $\varphi = 0$, while the second, negative-tension standard model brane is placed at $\varphi = \pi$. In this framework, gravity originates on the Planck brane and the graviton wave function is exponentially suppressed away from the brane along the fifth dimension due to the warp factor. Consequently, the $O(M_{\rm Pl})$ operators generated on the Planck brane yield low-energy effects on the standard model brane with a typical scale of $\Lambda_\pi = \overline{M}_{\rm Pl}\exp(-\pi k R_c)$, where $\overline{M}_{\rm Pl} \equiv M_{\rm Pl}/\sqrt{8\pi}$ is the reduced Planck mass. The hierarchy between the Planck and electroweak scales is solved if $\Lambda_\pi$ is $\sim 1$~TeV, which can be achieved with little fine tuning by requiring $kR_c \approx 12$. It has been shown~\cite{Wise} that the size of the extra dimension can be stabilized by the presence of a bulk scalar field (the radion). Consequently, the hierarchy problem in the Randall-Sundrum model is solved naturally for $k \sim 10/R_c \sim (10^{-2} - 10^{-1})\overline{M}_{\rm Pl}$, since $\overline{M}_{\rm Pl}$ is the only fundamental scale in this model and both $k$ and $1/R_c$ are of the same order as this scale. 

It is convenient to introduce a dimensionless parameter $\tilde{k} \equiv k/\overline{M}_{\rm Pl}$, which defines the strength of coupling between the graviton and the standard model fields. Theoretically preferred value of $\tilde{k}$ is between 0.01 and 0.1. For larger coupling theory becomes non-perturbative; if the coupling is too small, an undesirably high amount of fine tuning is still required to solve the hierarchy problem.

\begin{figure}[tb]
\epsfxsize=4.2in
\hfill\epsffile{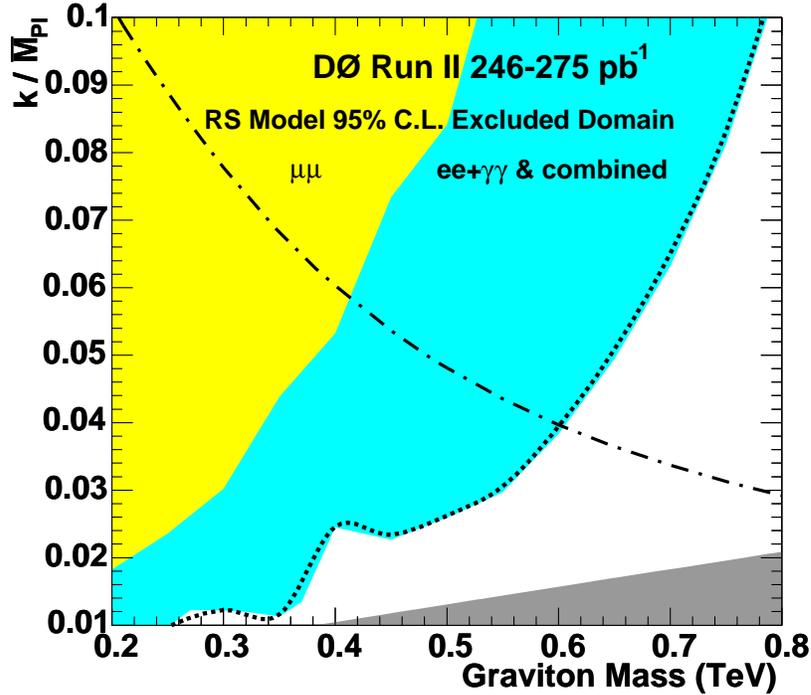}
\caption{Limits on the Randall-Sundrum model parameters $M_1$ and $\tilde{k} =\kMPl$ from the \D0 experiment~\protect\cite{RSD0}. The light-shaded area has been excluded in the dimuon channel; the medium-shaded area shows the exclusion obtained in the dielectron and diphoton channels; the dotted line corresponds to the combination of all three channels. The area below the dashed-dotted line is excluded from the precision electroweak data (see~\protect\cite{RSext}). The dark shaded area in the lower right-hand corner corresponds to $\Lambda_\pi > 10$ TeV, which requires undesirably high amount of fine tuning.}
\label{fig:RS}
\end{figure}

In the simplest form of the Randall-Sundrum model only gravitons are allowed to propagate in the fifth dimension; in the extensions of the model other particles are allowed to ``live'' in the bulk~\cite{RSext}. In both cases, since gravitons are allowed to propagate in the bulk, from the point of view of a 4D-observer on the standard model brane, they acquire Kaluza-Klein modes with the masses given by the subsequent zeroes $x_i$ of the Bessel function $J_1$ ($J_1(x_i) = 0$): $M_i = k x_i e^{-\pi k R_c} = \tilde{k} x_i \Lambda_\pi$. The mass of the first excited mode is then $M_1 \approx 3.83\tilde{k}\Lambda_\pi \sim 1$~TeV. The zeroth Kaluza-Klein mode of the graviton remains massless and couples to the standard model fields with the gravitational strength, $1/M_{\rm Pl}$, while the excited graviton modes couple with the strength of $1/\Lambda_\pi$. Coupling $\tilde{k}$ governs both the graviton width and the production cross section at colliders (both are proportional to $\tilde{k}^2$).

Indirect limits on the Randall-Sundrum model parameters come from precision electroweak data (dominated by the constraint on the $S$ parameter)~\cite{RSext}. The only dedicated search for Randall-Sundrum gravitons at colliders so far has been accomplished by the \D0 experiment at the Fermilab Tevatron and set direct limits~\cite{RSD0} on $\tilde{k}$ as a function of the mass of the first Kaluza-Klein mode of graviton, $M_1$. Both direct and indirect limits are summarized in figure~\ref{fig:RS}. As seen from the figure, for masses $M_1 \approx 3.83\tilde{k}\Lambda_\pi \gtrsim 800$~GeV most of theoretically preferred range for $\tilde{k}$ is still allowed.

\section{Astronomical black holes}

While very few scientists doubt that black holes exist somewhere in the universe, and perhaps are even abundant, all the astronomical black hole candidates discovered so far are only found via indirect means. There are several ways astronomers look for black holes. For example, one of the first established black hole candidates, Cygnus X-1 binary system, was found by observing the orbital periods in superbright binary systems. The presence of a black hole as one of the two stars in Cygnus X-1 was inferred from the large total power dissipated by the system, large estimated mass of the unseen companion star, and from extremely short time-scale of the intensity variations. Similar arguments lead astronomers to believe that quasars are powered by massive black holes. Furthermore, by observing X-ray flares in the active galactic nuclei, it is speculated that they are caused by large objects falling inside the nuclei, being attracted by supermassive ($\sim 10^6$ solar masses) black holes located in their centers. The presence of these supermassive black holes is also inferred by the motion of stars near the galactic centers. There is even an evidence for such a supermassive black hole in the center of our own galaxy, the Milky Way. Clearly, these observations prove the presence of massive compact objects in many of the binary systems or galaxies; however, proving that the critical density, necessary for a gravitational collapse into a black hole, has been achieved in these systems is rather complicated. The recent announcement of an observation of a ``strange'' star, apparently more dense than a neutron star~\cite{strangestar}, given the lack of a compelling cosmological explanation or even motivation for such an object, might indicate that there is a problem with the current methods of density estimates in extremely remote systems.

Even the direct dynamical observations only probe the region of space at least two orders of magnitude further away from the center of a black hole candidate than the estimated radius of its event horizon (often referred to as the Schwarzschild radius, $R_S$). To date, the closest we have been able to probe our own galactic center is $\approx 600 R_S$~\cite{Ghez}.

Other phenomena predicted for black holes in general relativity, such as frame dragging, have been observed as well. However, none of the existing black hole candidates have been tagged by several independent means so far. Given the large number of objects studied by the astronomers in their quest for black holes, it is questionable how unambiguous the single tags are.

Unfortunately, the most prominent feature of a black hole --- its Hawking radiation~\cite{Hawking}~--- has not been observed yet and is very unlikely to be ever observed by astronomical means. Indeed, even the smallest (and therefore the hottest) astronomical black holes with the mass close to the Oppenheimer-Volkov limit~\cite{Oppenheimer} of $\sim 3$ solar masses, have Hawking temperatures of only $\sim 100$~nK, which corresponds to the wavelength of Hawking radiation of $\sim 100$~km, and the total dissipating power of puny $\sim 10^{-28}$~W.\footnote{Note that the event horizon temperature of these black holes is much lower than the temperature of the cosmic microwave background radiation, so at the present time the black holes are growing due to the accretion of relic radiation much faster than they are evaporating. The evaporation process will take over only when the expanding universe cools down below the event horizon temperature.}

Not only the event horizon of these black holes is nearly as cold as the lowest temperature ever achieved in the lab (the 1997 Nobel prize in physics~\cite{cooling} was given to Chu, Cohen-Tannoudji, and Phillips who reached the temperature as low as $\sim 1$~$\mu$K via optical cooling), but the wavelength of the radiation dissipated by a black hole is far from the visible spectrum and is resemblant of that of an AM radio station. Trying to detect such a radio broadcast with a vanishing transmitting power from thousands of light-years away is but impossible. Given that the black hole dissipating power corresponds to only $\sim 100$ photons per second emitted by its entire event horizon and that the closest known black hole candidate is still over a thousand of light years away from us, not a single Hawking radiation photon ever hit our Earth since it has been formed! (In fact, one would have to wait $\sim 10^{14}$ years to observe a single photon from such a black hole to hit the Earth.) Thus, if the astronomical black holes were the only ones to exist, the Hawking radiation would be always just a theoretical concept, never testable experimentally.

While Hawking radiation would constitute a definitive proof of the black hole nature of a compact object, there are other, indirect means of identifying the existence of the event horizon around it. It has been suggested~\cite{Narayan} that the lack of Type I X-ray bursts in the binary systems identified as black hole candidates, implies the presence of the event horizon. The argument is based on the fact that such X-ray bursts are frequent in the neutron star binaries, similar in size and magnitude to the black hole candidates. Consequently, if the black hole candidates did not have the event horizon, one would expect to see a similar X-ray burst activity, which contradicts the observations. An analogous argument applies to the X-ray supernovae that are believed to contain black holes. These supernovae are much dimmer than similar supernovae that are believed to contain neutron stars, which is considered to be an evidence for the formation of an event horizon around the black hole candidates. Unfortunately, an evidence based on a non-observation of a particular predicted phenomenon is inherently much more model-dependent and circumstantial than the one based on an observation of a certain effect. Perhaps, a more promising way to prove the existence of the event horizon around some of the black hole candidates is to compare the accretion disk shapes in various X-ray binaries~\cite{Narayan1}. Black hole binaries are expected to have advection-dominated accretion flows, drastically different from thin accretion disks, typical of subcritical binaries.

Probably, the best evidence for the existence of astronomical black holes would come from an observation of gravitational waves created in the collisions of two black holes, which LIGO and VIRGO detectors are looking for. However, current sensitivity of these interferometers is still short of the expected signal, even in optimistic cosmological scenarios.

This leads us to other places to look for black holes that are much smaller and consequently much hotter and easier to detect than their astronomical counterparts.

\section{Mini black holes in large extra dimensions}

As was pointed out several years ago~\cite{adm,bf,ehm}, an exciting consequence of low-scale quantum gravity is the possibility of producing black holes (BH) at particle accelerators. More recently, this phenomenon has been quantified~\cite{dl,gt}for the case of particle collisions at Large Hadron Collider (LHC), resulting in a mesmerizing and unexpected prediction that future colliders would produce mini black holes at enormous rates (e.g., $\sim 1$~Hz at the LHC for $\MP = 1$~TeV), thus qualifying for black-hole factories. With the citation index of the original papers~\cite{dl,gt} well over three hundred, the production of mini black holes in the lab became one of the most actively studied and rapidly evolving subjects in the phenomenology of models with low-scale gravity.

In general relativity, black holes are well understood if their mass \mbh far exceeds the Planck scale. Consequently, in the model with large extra dimensions general relativity would give accurate description of black-hole properties when its mass is much greater than the fundamental (multidimensional) Planck scale
$\MP \sim 1$~TeV. As its mass decreases and approaches \MP, the black hole becomes a quantum gravity object with unknown and presumably complex properties. 

In this section, following \cite{dl}, we will ignore this obstacle\footnote{Some of the properties of the ``stringy'' subplanckian ``precursors'' of black holes are discussed in \protect\cite{de} and later in this review.} and estimate the properties of light black holes by simple semiclassical arguments, strictly valid only for $\MBH \gg \MP$. We expect this to be an adequate approximation, since the important experimental signatures rely on two simple qualitative properties: the absence of small couplings and the ``democratic" nature of black hole decays, both of which may survive as average properties of the light descendants of black holes. 

As we expect unknown quantum gravity effects to play an increasingly important role for the black hole mass approaching the fundamental Planck scale, following the prescription of~\cite{dl}, we do not consider black hole masses below the Planck scale. It is expected that the black-hole production rapidly turns on, once the relevant energy threshold $\sim\! \MP$ is crossed. At lower energies, we expect black-hole production to be exponentially suppressed due to the string excitations or other quantum effects. 

We will first focus on the production in particle collisions and subsequent decay of small Schwarzschild black holes with the size much less than the compactification radius of extra dimensions. In this case, standard Schwarzschild solution found for a flat $(3+n)$-dimensional metric fully applies. The expression for the Schwarzschild radius $R_S$ of such a black hole in $(3+n)$ spacial dimensions is well known~\cite{mp}:
\begin{equation}
   R_S(\MBH) = \frac{1}{\sqrt{\pi}\MP}\left[\frac{\MBH}{\MP}
   \frac{8\Gamma\left(\frac{n+3}{2}\right)}{n+2}\right]^\frac{1}{n+1}.
\label{eq:RS}
\end{equation}
Given $\MP \sim 1$~TeV and taking into account the fact that black hole masses accessible at the next generation of particle colliders and in ultrahigh-energy cosmic ray collisions are at most a few TeV, we note that the Schwarzschild radius of such black holes is $\sim 1/\MP$, i.e. indeed much smaller than the size of large extra dimensions even when their number approaches six or seven (the preferred number of extra dimensions expected in string theory). We also note that for $\MBH \sim \MP$ the Schwarzschild radius does not depend significantly on the number of extra dimensions $n$.

Given the current lower constraints on the fundamental Planck scale in the model with large extra dimensions of $\approx 1$~TeV~\cite{collider}, the black holes that we may be able to study at colliders and in cosmic rays will be barely transplanckian. Hence, the unknown quantum corrections to their classical properties are expected to be large, and therefore it is reasonable to focus only on the most robust properties of these mini black holes that are expected to be affected the least by unknown quantum gravity corrections. Consequently, when discussing production and decay of black holes, we do not consider the effects of spin and other black hole quantum numbers, as well as grey-body factors. Further in the review, we discuss some of the subsequent attempts to take these effects into account using semiclassical approximation.

\subsection{Black-hole production in particle collisions}

Consider two partons with the center-of-mass energy $\sqrt{\hat s} =
\MBH$ colliding head-on. Semiclassical reasoning
suggests that if the impact parameter of the collision is less than the Schwarzschild radius $R_S$, corresponding to this energy, a black hole with the mass \mbh is formed. Therefore the total cross section of black hole production in particle collisions can be estimated from pure geometrical arguments and is of order $\pi R_S^2$.

Soon after the original calculations~\cite{dl,gt} have appeared, it has been suggested~\cite{Voloshin} that the geometrical cross section is in fact exponentially suppressed, based on the Gibbons-Hawking action~\cite{gh} argument. Detailed subsequent studies performed in simple string theory models~\cite{de}, using full general relativity calculations~\cite{GRcollisions}, or a path integral approach~\cite{path} did not confirm this finding and proved that the geometrical cross section is modified only by a numeric factor of order one. A flaw in the Gibbons-Hawking action argument of \cite{Voloshin} was further found in \cite{jt}: the use of this action implies that the black hole has been already formed, so describing the evolution of the two colliding particles before they cross the event horizon and form the black hole via Gibbons-Hawking action is not justified. By now there is a broad agreement that the production cross section is not significantly suppressed compared to a simple geometrical approximation, which we will consequently use through this review.

\begin{figure}[tbp]
\epsfxsize=4.2in
\hfill\epsffile{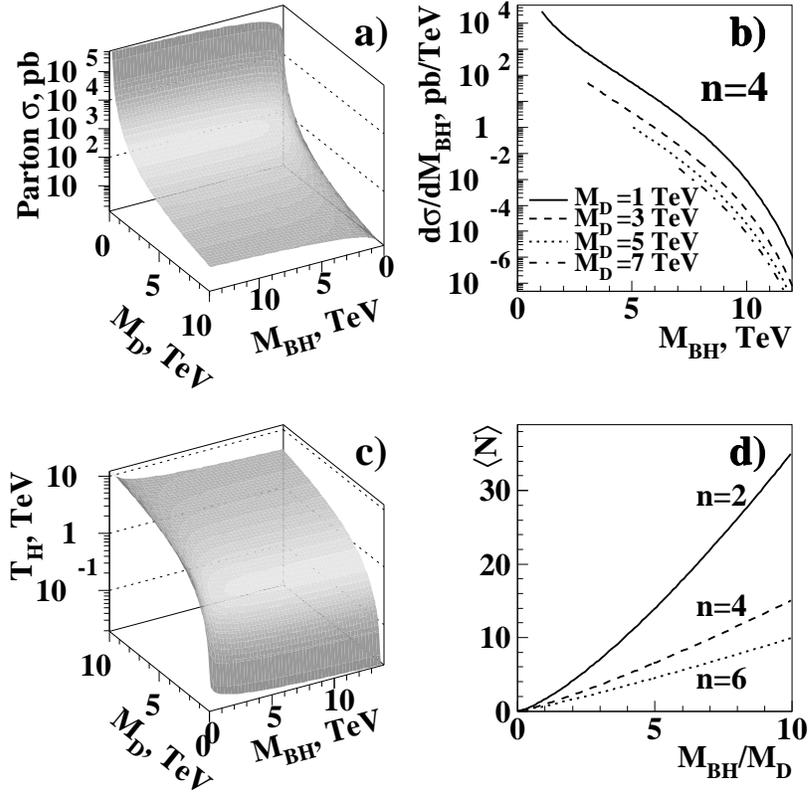}
\caption{a) Parton-level and b) differential production cross section of black holes in $pp$ collisions at the LHC. Also shown: c) Hawking temperature and d) average particle multiplicity in black-hole decays. Adapted from~\cite{dl}.}
\label{fig:rsth}
\end{figure}

Using the expression (\ref{eq:RS}) for the Schwarzschild radius~\cite{mp}, 
we derive the following parton level black hole production cross section~\cite{dl}:
\begin{equation}
    \sigma(\MBH) \approx \pi R_S^2 = \frac{1}{\MP^2}
    \left[
      \frac{\MBH}{\MP} 
      \left( 
        \frac{8\Gamma\left(\frac{n+3}{2}\right)}{n+2}
      \right)
    \right]^\frac{2}{n+1}.
\label{eg:sigma}
\end{equation}
This cross section, as a function of the fundamental Planck scale and the black hole mass, is shown in figure~\ref{fig:rsth}a.

In order to obtain the production cross section in $pp$ collisions at the LHC, 
we use the parton luminosity approach~\cite{dl,gt,EHLQ}:
$$
    \frac{d\sigma(pp \to \mbox{BH} + X)}{d\MBH} = 
    \frac{dL}{dM_{\rm BH}} \hat{\sigma}(ab \to \mbox{BH})
    \left|_{\hat{s}=M^2_{\rm BH}}\right.,
$$
where the parton luminosity $dL/d\MBH$ is defined as the sum over
all the types of initial partons:
$$
    \frac{dL}{dM_{\rm BH}} = \frac{2\MBH}{s} 
    \sum_{a,b} \int_{M^2_{\rm BH}/s}^1  
    \frac{dx_a}{x_a} f_a(x_a) f_b(\frac{M^2_{\rm BH}}{s x_a}),
$$
and $f_i(x_i)$ are the parton distribution functions (PDFs). We
used the MRSD$-'$~\cite{MRSD} PDFs with the $Q^2$ scale taken
to be equal to \MBH, which is within the allowed range of these
PDFs for up to the kinematic limit at the LHC. The dependence of 
the cross section on the choice of PDF is only $\sim 10\%$.
The total production cross section for $\MBH > \MP$ at the LHC, 
obtained from the above equation, ranges between 15 nb and 1 pb for 
the Planck scale between 1 TeV and 5 TeV, and varies by $\sim 10\%$ 
for $n$ between 2 and 7. The differential cross section is shown in figure~\ref{fig:rsth}b.

\subsection{Black-hole evaporation}

In general relativity, black-hole evaporation is expected to occur in three distinct stages: ``balding,'' spin-down, and Hawking evaporation. During the first stage, the black hole loses its multipole momenta and quantum numbers via emission of gauge bosons until it reaches the Kerr solution for a spinning black hole; at the second stage it gets rid of the residual angular momentum and becomes a Schwarzschild black hole; and at the last stage it decays via emission of black-body radiation~\cite{Hawking} with a characteristic Hawking temperature:
\begin{equation}
    T_H = \MP
    \left(
      \frac{\MP}{\MBH}\frac{n+2}{8\Gamma\left(\frac{n+3}{2}\right)}
    \right)^\frac{1}{n+1}\frac{n+1}{4\sqrt{\pi}} = \frac{n+1}{4\pi R_S}
\label{eq:TH}
\end{equation}
of $\sim 1$~TeV~\cite{dl,gt}. The dependence of the Hawking temperature on the fundamental Planck scale and the black-hole mass is shown in figure~\ref{fig:rsth}c.

Note that if a certain quantum number (e.g., $B - L$) is gauged, it will be conserved in the process of black hole evaporation. Since the majority of black holes at the LHC are produced in quark-quark collisions, one would expect many of them to have the baryon number and fractional electric charge. Consequently, the details of black-hole evaporation process will allow to determine if these quantum numbers are truly conserved.

In quantum gravity, it is expected that there is a fourth, Planckian stage of black hole evaporation, which is reached when the mass of the evaporating black hole approaches the Planck scale. The details of the Planckian stage are completely unknown, as they are governed by the effects of quantum gravity, which should be dominant at such low black hole masses. Some authors speculate that the Planckian stage terminates with a formation of a stable or semi-stable black hole remnant with the mass $\sim M_{\rm Pl}$. Others argue that the evaporation proceeds until the entire mass of the black hole is radiated. The truth is that no predictions about the Planckian regime are possible, given our lack of knowledge of quantum gravity.

The average multiplicity of particles produced in the process of
black-hole evaporation is given by: $\langle N \rangle = \left\langle
\frac{\MBH}{E} \right\rangle$, where $E$ is the energy spectrum
of the decay products. In order to find $\langle N \rangle$, we note
that evaporation is a black-body radiation process, with the 
energy flux per unit of time given by Planck's formula:
\begin{equation}
	\frac{df}{dx} \sim \frac{x^3}{e^x + c},
\label{eq:flux}
\end{equation}
where $x \equiv E/T_H$, and $c$ is a constant, which depends on the quantum statistics of the decay products ($c = -1$ for bosons, $+$1 for fermions, and 0 for
Boltzmann statistics).

The spectrum of the decay products in the massless particle approximation is given by: $\frac{dN}{dE} \sim
\frac{1}{E}\frac{df}{dE} \sim \frac{x^2}{e^x + c}$. For averaging
the multiplicity, we use the average of the distribution in the 
inverse particle energy:
\begin{equation}
    \left\langle \frac{1}{E} \right\rangle =
    \frac{1}{T_H}\frac{\int_0^\infty dx \frac{1}{x}
    \frac{x^2}{e^x + c}}{\int_0^\infty dx\frac{x^2}{e^x + c}}
    = a/T_H,
\label{eq:eav}
\end{equation}
where $a$ is a dimensionless constant that depends on the type of
produced particles and numerically equals 0.68 for bosons, 0.46
for fermions, and $\frac{1}{2}$ for Boltzmann statistics. Since a
mixture of fermions and bosons is produced in the black hole decay, we can
approximate the average by using Boltzmann statistics, which gives
the following formula for the average multiplicity: $\langle N
\rangle \approx \frac{\MBH}{2T_H}$. Using expression (\ref{eq:TH}) for
Hawking temperature, we obtain:
\begin{equation}
    \langle N \rangle = \frac{2\sqrt{\pi}}{n+1}
    \left(\frac{\MBH}{\MP}\right)^\frac{n+2}{n+1}
    \left(\frac{8\Gamma\left(\frac{n+3}{2}\right)}{n+2}\right)^\frac{1}{n+1},
\label{eq:nav}
\end{equation}
which corresponds to about half-a-dozen for typical black hole masses accessible 
at the LHC, see figure~\ref{fig:rsth}d. 

Na\"ively, one would expect that a large fraction of Hawking radiation is emitted in the form of gravitons, escaping in the bulk space. However, as was shown in~\cite{ehm}, this is not the case, since the wavelength $\lambda = {2 \pi \over T_H}$
corresponding to the Hawking temperature is larger than the size of the black hole. Therefore, the black hole acts as a point-radiator and consequently emits mostly $s$-waves. Since the $s$-wave emission is sensitive only to the radial coordinate, bulk radiation per graviton degree of freedom is the same as radiation of any standard model degree of freedom on the brane. While many angular degrees of freedom are available in the bulk space, the $s$-wave emission cannot take advantage of them, thus suppressing bulk graviton component. Since there are many more particles on 
the brane than in the bulk space, this has the crucial consequence that the black hole mainly decays to visible standard model particles.

Since the gravitational coupling is flavour-blind, a black hole emits all the 
$\approx 120$ standard model particle and antiparticle degrees of freedom with 
roughly equal probability. Accounting for the colour and spin and ignoring the graviton emission, we expect 
$\approx 75\%$ of particles produced in black hole decays to be quarks and gluons, 
$\approx 10\%$ charged leptons, $\approx 5\%$ neutrinos, and $\approx 5\%$ 
photons or $W/Z$ bosons, each carrying hundreds of GeV of energy. 
Similarly, if there exist new particles with masses $\lesssim 100$~GeV, 
they would be produced in the decays of black holes with the probability 
similar to that for the standard model species. For example, a sufficiently light 
Higgs boson is expected to be emitted with $\sim 1\%$ probability. This has exciting consequences for searches for new physics at the LHC and beyond, as the production cross section for any new particle via this mechanism is large and depends only weakly on the particle mass, in contrast with an exponential dependence characteristic of direct production.

A relatively large fraction of prompt and energetic photons, electrons, 
and muons expected in the high-multiplicity black hole decays would 
make it possible to select pure samples of black holes, which are also 
easy to trigger on~\cite{dl,gt}. At the same time, only a small fraction 
of particles produced in the black hole decays are undetectable gravitons 
and neutrinos, so most of the black hole mass is radiated in the form of visible 
energy, making it easy to detect.

It has been argued~\cite{chromosphere} that the fragmentation of quarks and jets emitted in the black hole evaporation might be significantly altered by the presence of a dense and hot QCD plasma (``chromosphere'') around the event horizon. If this argument is correct, one would expect much softer hadronic component in the black hole events. However, we would like to point out that one would still have a significant number of energetic jets due to the decay of weakly interacting $W/Z$ and Higgs bosons, as well as tau leptons, emitted in the process of black hole evaporation and penetrating the chromosphere before decaying into jetty final states. In any case, tagging of the black hole events by the presence of an energetic lepton or a photon and large total energy deposited in the detector is a fairly model-independent approach.

\begin{figure}[tbp]
\epsfxsize=4.2in
\hfill\epsffile{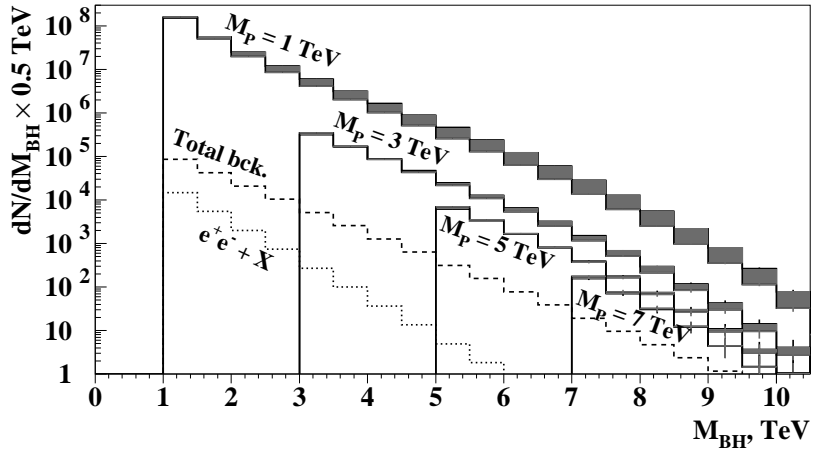}
\caption{Number of black holes produced at the LHC in the electron or photon decay 
channels (assuming 15\% efficiency for an electron or a photon tag), with 100~\protect\ifb of integrated luminosity, as a function of the black hole 
mass. The shaded regions correspond to the variation in the number of events 
for $n$ between 2 and 7. The dashed line shows total standard model background 
(from inclusive $Z(ee)$ and direct photon production). The dotted line 
corresponds  to the $Z(ee)+X$ background alone. Adapted from \protect\cite{dl}.}
\label{fig:nbh}
\end{figure}

In figure~\ref{fig:nbh} we show the number of black hole events tagged by the presence of an energetic electron or photon among the decay products in 100~fb$^{-1}$ of data collected at the LHC, along with estimated backgrounds, as a function of the black hole mass~\cite{dl}. It is clear that very clean and large samples of black holes can be produced at the LHC up to the Planck scale of $\sim 5$ TeV. Note that the black hole discovery potential at the LHC is maximized in the $e/\mu+X$ channels, where background is much smaller than that in the $\gamma+X$ channel (see figure~\ref{fig:nbh}). The reach of a simple counting experiment extends up to $\MP \approx 9$ TeV ($n=2$--7), for which one would expect to see a handful of black hole events with negligible 
background. 

The lifetime of a black hole can be estimated using Stefan's law of thermal radiation. Since black hole evaporation occurs primarily in three spatial dimensions, the canonical 3-dimensional Stefan's law applies, and therefore the power dissipated by the Hawking radiation per unit area of the event horizon is $p = \sigma T_H^4$, where $\sigma$ is the Stefan-Boltzmann constant and $T_H$ is the Hawking temperature. Since the effective evaporation area of a black hole is the area of a 3D-sphere with radius $R_S$ and Stefan's constant in natural units is $\sigma = \pi^2/60 \sim 1$, dropping numeric factors of order unity we obtain the following expression for the
total power dissipated by a black hole: $P \sim R_S^2 T_H^4 \sim R_S^{-2}$.

The black hole lifetime $\tau$ then can be estimated as: $\tau \sim \MBH/P \sim \mbh R_S^2,$ and using (\ref{eq:RS}) we find:
\begin{equation}
  \tau \sim \frac{1}{\MP}\left(\frac{\mbh}{\mp}\right)^\frac{n+3}{n+1}.
\label{eq:tau}
\end{equation}
Therefore, a typical lifetime of a mini black hole is $\sim 10^{-27} - 10^{-26}$ s. A multi-TeV black hole would have a relatively narrow width $\sim 100$~GeV, i.e. similar to, e.g., a $W'$ or $Z'$ resonance of a similar mass. This is not surprising, as the strength of gravity governing the black hole evaporation rate is similar in the model with large extra dimensions to that of electroweak force responsible for the $W'$ or $Z'$ decay rates.

\subsection{Accounting for the black hole angular momentum and grey-body factors}

In the above discussion we used a number of approximations in deriving production cross section and decay properties of mini black holes. While reliable accounting for more complicated effects related to quantum properties of black holes (spin, quantum numbers, etc.) is not possible without intimate knowledge of the underlying theory of quantum gravity, a number of authors attempted to estimate some of the above effects using simple semiclassical approach. While we don't believe that these estimates are any more reliable than the ones obtained in the above simple approximation, we discuss these refinements here in some detail.

The most studied properties of mini black holes beyond our simple picture are the effects of its angular momentum and grey-body factors, which have to do with the emissivity of particles of various types in the process of black hole evaporation.

An emissivity of a certain type of particle depends, in general, on its spin, $S$. Indeed, for a small black hole the Schwarzschild radius is comparable with the Compton wavelength of the emitted particles. Thus the wave function of a spin-0 particle would have more overlap with the black hole event horizon than that for a spin-1/2 or spin-1 particle. Consequently, one would expect that scalar particles are emitted by a black hole more efficiently than spin-1/2 or higher-spin particles. This qualitative feature can be parameterized as a grey-body factor, $g(S,x)$, where $x = E/T_H$ is a dimensionless variable proportional to the energy of the particle. Thus, the modified expression (\ref{eq:flux}) for the emitted flux is:
\begin{equation}
	\frac{df}{dx} \sim \frac{g(S,x) x^3}{e^x + c}.
\end{equation}

The grey-body factors can be calculated classically in general relativity. This calculation has been extended recently to the case of multidimensional black holes in the model with large extra dimensions by a number of authors~\cite{grey-body,grey-body1,grey-body2}.

\begin{figure}[htbp]
\epsfxsize=4.2in
\hfill\epsffile{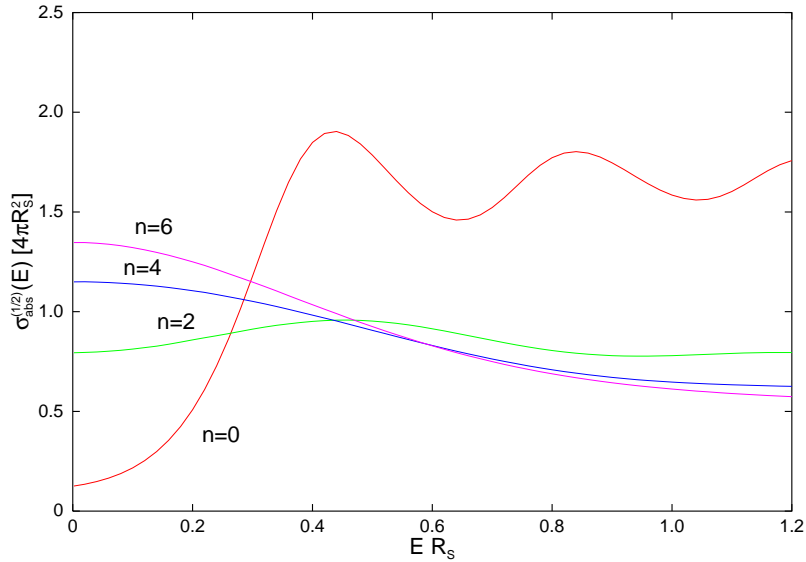}
\caption{Grey-body factors for the emissivity of spin-$1/2$ particles on the brane as a function of their energy, $E$. Vertical axis shows the effective area of the black hole horizon for the emission of a spin-$1/2$ particle expressed in the units of the geometrical surface of the event horizon of a three-dimensional black hole, $4\pi R_S^2$. Adapted from~\protect\cite{grey-body1}.}
\label{fig:GB}
\end{figure}

In special cases, the grey-body factors can be calculated analytically, but most general calculations that exist to date are performed numerically. The grey-body factor for the emissivity of spin-$1/2$ particles as a function of their energy is given in figure~\ref{fig:GB} from~\cite{grey-body1}. As seen from the figure, the effect of grey-body factors is small for characteristic black-body radiation energies and $n \ge 2$.

Recently, grey-body factors for the brane and bulk emission of gravitons have been calculated as well~\cite{grey-body2}. While complete calculations applicable to all graviton energies are not available yet, it appears that the fraction of Hawking radiation of bulk and brane gravitons is relatively small, except for the cases of very high black-hole angular momentum or very large number of extra dimensions ($n = 7$).

Since black holes in particle collisions are produced by particles with non-vanishing impact parameter, they may carry non-zero angular momentum and thus produce a spinning black hole. Given that the characteristic impact parameter is $R_S$, while the relative momentum of the colliding particles is $\MBH$, one expects a black hole produced in such a grazing collision to carry angular momentum 
$$
	L \sim R_S \MBH \sim \left(\frac{\MBH}{\MP}\right)^\frac{n+2}{n+1}.
$$
For black holes with masses close to the fundamental Planck scale, typical angular momentum is of the order of one. Note that although numerically this is rather small angular momentum, it is close to the maximum angular momentum that a black hole of such a small mass could have.

For a rotating black hole, the solution of Einstein's equation is given by Kerr's, rather than Schwarzschild's formula and contains explicit dependence on the angular momentum of the black hole. A number of authors have studied properties of Kerr black holes in theories with extra dimensions~\cite{BH-spin}. Both the spin-down effects and modification of the grey-body factors due to the angular momentum of the black hole have been looked at. In general, for a large number of extra dimensions and for a large initial angular momentum of the black hole it is expected that modification of the classical black-body radiation picture becomes non-trivial and will have to be taken into account for an accurate description of black-hole evaporation.

Several authors looked at other details of Hawking evaporation process. The effects of using the microcanonical ensemble approach, which takes into account that the energy of the emitted particles is comparable to the black-hole mass, have been discussed in \cite{microcanonical} and generally result in the increased black-hole lifetime. The recoil effect in the evaporation has been studied~\cite{recoil} as well. 

To summarize, while the recent angular momentum and grey-body factor results are important and encouraging, for the black holes with the masses close to the fundamental Planck scale they are likely to be modified in a profound way via unknown quantum corrections. Thus detailed studies of the particle content in black-hole evaporation probably won't be possible until either their discovery at the LHC or a formulation of complete theory of quantum gravity.

\subsection{Other black hole signatures and studies}

While the detection of Hawking radiation from rapidly evaporating TeV-scale black holes remains one of the most clear and well-studied signatures at the LHC, several authors studied complementary signatures as well. Among those most notable are studies of the suppression of dijet production at the invariant masses exceeding the fundamental Planck scale~\cite{bf,dijets}, anomalous inclusive hadron production due to products of black hole decays~\cite{hadrons}, and production of black holes in heavy ion collisions at the LHC~\cite{HI}.

Recently, there have been attempts to estimate the decrease of the black-hole production cross section due to inelasticity in parton collisions. The reason for the inelasticity is the emission of gravitational waves during the formation of black holes, some of which may not be trapped within the event horizon and escape. For discussion of this effect, see \cite{inelastic}.

\subsection{Testing Hawking radiation in decays of black holes}

\begin{figure}[htbp]
\epsfxsize=4.2in
\hfill\epsffile{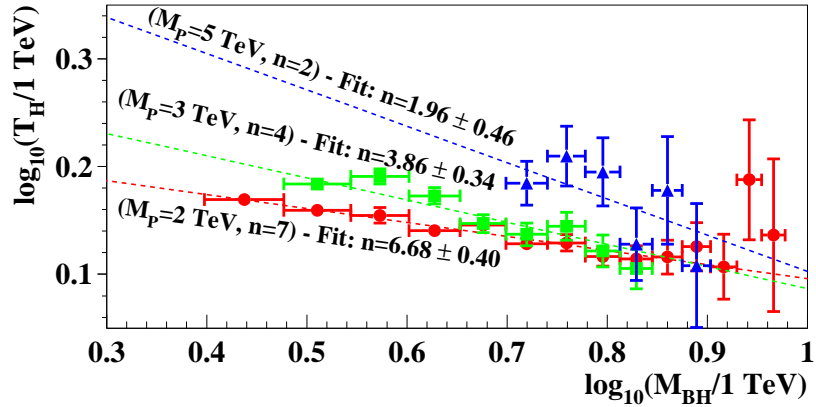}
\caption{Determination of the
dimensionality of  space via Wien's displacement law at the LHC with
100~\protect\ifb\ of data. Adapted from \protect\cite{dl}.}
\label{Wien}
\end{figure}

A sensitive test of properties of Hawking radiation can be performed by measuring the relationship between the mass of the black hole (reconstructed from the total energy of all the visible decay products) and its Hawking temperature (measured from the energy spectrum of the emitted particles). One can use the measured \mbh vs. \TH\ dependence to determine both the fundamental Planck scale \mp and the dimensionality of space $n$. This is a multidimensional equivalent of the Wien's law of thermal radiation. It is particularly interesting that the dimensionality of extra space can be determined in a largely model-independent way via taking a logarithm of both parts of (\ref{eq:TH}) for Hawking temperature: $\log_{10}(T_H/1 TeV) = -\,\frac{1}{n+1}\log_{10}(\MBH/1 TeV) + \mbox{const}$, where the constant does not depend on the black hole mass, but only on the \mp and on detailed properties of the bulk space, such as relative size of extra dimensions~\cite{dl}. Therefore, the slope of a straight-line fit to the $\log_{10}(T_H/1 TeV)$ vs. $\log_{10}(\MBH/1 TeV)$ data offers a direct way of determining the dimensionality of space. The reach of this method at the LHC is illustrated in figure~\ref{Wien} and discussed in detail in \cite{dl}. Note that the determination of the dimensionality of space by this method is fundamentally different from other ways of determining $n$, e.g. by studying a monojet signature or a virtual graviton exchange processes, also predicted in the models with large extra dimensions. The latter always depend on the volume of extra space, and therefore cannot provide a direct way of measuring $n$ without making assumptions about the relative size of large extra dimensions. The former depends only on the area of the event horizon of a black hole, which is not sensitive to the size of large extra dimensions.

An interesting possibility studied in \cite{de} is production of a precursor of a black hole, i.e. a long and jagged highly exited string state, dubbed as a ``string ball'' due to its folding in a ball-like object via a random walk. As shown in \cite{de}, there are three characteristic string ball production regimes, which depend on the mass of the string ball $M$, the string scale $M_S < \MP$,  and the string coupling $g_s < 1$. For $M_S < M < M_S/g_s$, the production cross section increases $\propto M^2$, until it reaches saturation at $M \sim M_S/g_s$ and stays the same up to the string ball mass $\sim M_S/g_s^2$, at which point a black hole is formed and the production cross section agrees with that from \cite{dl}.

A string ball has properties similar to those of a black hole, except that its evaporation temperature, known as Hagedorn temperature~\cite{Hagedorn}, is constant: $T_S = M_S/(2\sqrt{2}\pi)$. Thus, the correlation between the temperature of the characteristic spectrum and the string ball mass may reveal the transition from the Hagedorn to Hawking regime, which can be used to estimate $M_S$ and $g_s$. Another possibility is a production of higher-dimensional objects, e.g. black $p$-branes, rather than spherically symmetric black holes ($p=0$)~\cite{blackbranes}. For a detailed review see, e.g. \cite{kingman}.

\subsection{Simulation of black hole production and decay}

In order to study properties of black holes at colliders, it's important to have tools capable of simulating their production and decay. Currently, there are two Monte Carlo generators capable of doing this: TRUENOIR~\cite{truenoir} and CHARYBDIS~\cite{charybdis}.

The TRUENOIR generator~\cite{truenoir}, available since 2001, is a plug-in module for the PYTHIA~\cite{PYTHIA} Monte Carlo package. It simulates the production and decay of black holes as described in \cite{dl}, assuming conservation of the individual baryon and lepton numbers, as well as the electric and colour charges. It does not include any grey-body factors for the black-hole decay and assumes that the evaporation takes place in a single, Schwarzschild phase, thus ignoring any spin effects. While the deficiency of these assumptions is obvious, the logic behind this simple approach is that given the uncertainties related to unknown quantum corrections, an incremental improvement from taking into account classically calculated corrections is minor.

The CHARYBDIS generator~\cite{charybdis} is available since 2003 and is interfaced with the HERWIG~\cite{HERWIG} Monte Carlo package. The authors of CHARYBDIS incorporated classically calculated grey-body factors and also gave the user flexibility of defining when the Schwarzschild stage terminates and a stable Planckian remnant of the black hole is formed.

\begin{figure}[tbp]
\begin{flushright}
\hspace*{2.0in}{\large\bf (a)}\hspace{1.9in}{\large\bf (b)}
\vskip -0.03in
\epsfxsize=2.2in
\epsffile{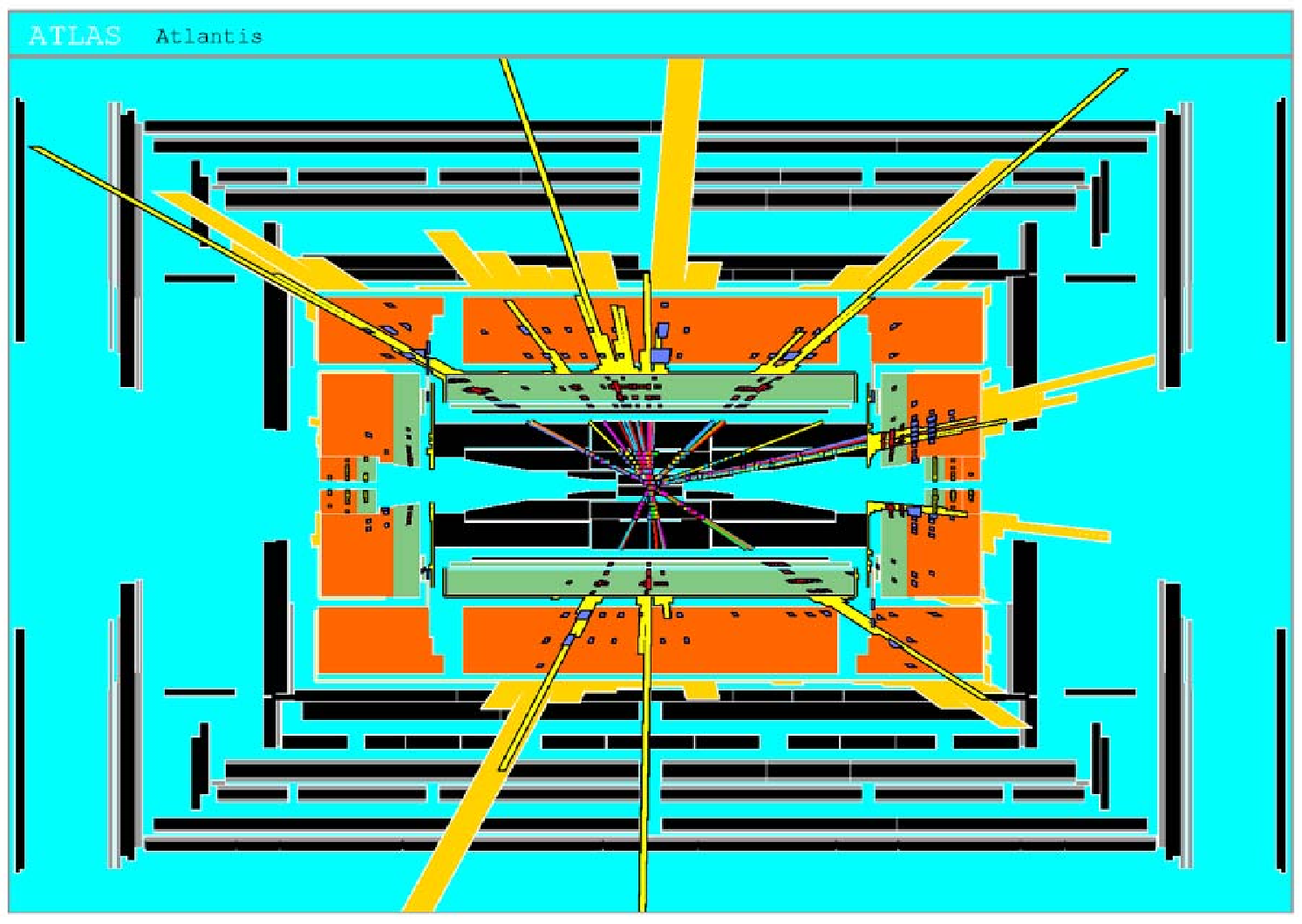}
\epsfxsize=1.9in
\epsffile{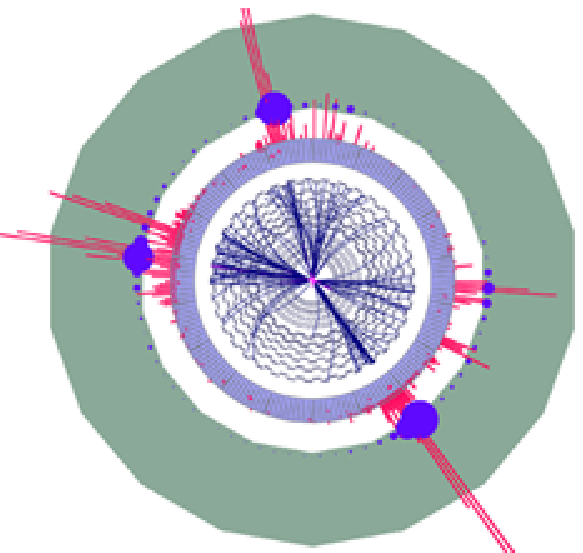}
\caption{(a) A typical black hole event, generated with CHARYBDIS~\protect\cite{charybdis}, as seen by the ATLAS detector (courtesy of ATLAS-Japan group). (b) A typical black hole event, generated with TRUENOIR~\protect\cite{truenoir}, as seen by the CMS detector (courtesy of Albert De Roeck).}
\label{fig:events}
\end{flushright}
\end{figure}

Both generators have been successfully interfaced with the LHC detector simulation packages. Figure~\ref{fig:events} shows simulations of a black hole event in the ATLAS and CMS detectors, using the CHARYBDIS and TRUENOIR Monte Carlo packages, respectively. A number of dedicated studies~\cite{LHC-studies} of the black-hole decay signatures at the LHC have been performed using these Monte Carlo generators.

\subsection{Discovering new particles in black hole decays}

As was mentioned earlier, new particles with the mass $\sim 100$~GeV would be produced in the process of black hole evaporation with a relatively large probability: $\sim 1\%$ times the number of their quantum degrees of freedom. Consequently, it may be advantageous to look for new particles among the decay products of black holes in large samples accessible at the LHC and other future colliders.

As an example, following ~\cite{glBH}, we demonstrate the discovery potential of a black-hole sample to be collected at the LHC for a light Higgs boson. We pick the Higgs boson mass of 130 GeV, which is still allowed in low-scale supersymmetry models, but makes it quite hard to discover Higgs via direct means either at the Fermilab Tevatron~\cite{SUSYHiggs} or at the LHC. We consider the decay of the Higgs boson into a pair of jets (with the branching fraction of 67\%), dominated by the $b\bar b$ final state (57\%), with an additional 10\% contribution from the $c\bar c$, $gg$, and hadronic $\tau\tau$ final states.

We model the production and decay of the black holes with the TRUENOIR Monte Carlo 
generator~\cite{truenoir}. We used a 1\% probability to emit the Higgs particle in the black hole decay. We reconstruct final state particles within the acceptance of a typical LHC detector and smear their energies with the expected resolutions. We select the black hole events by requiring an energetic electron, photon, or a muon, as well as total multiplicity of energetic objects (jets, electrons, photons, or muons) of at least four.

\begin{figure}[tbp]
\begin{flushright}
\epsfxsize=4.2in
\epsffile{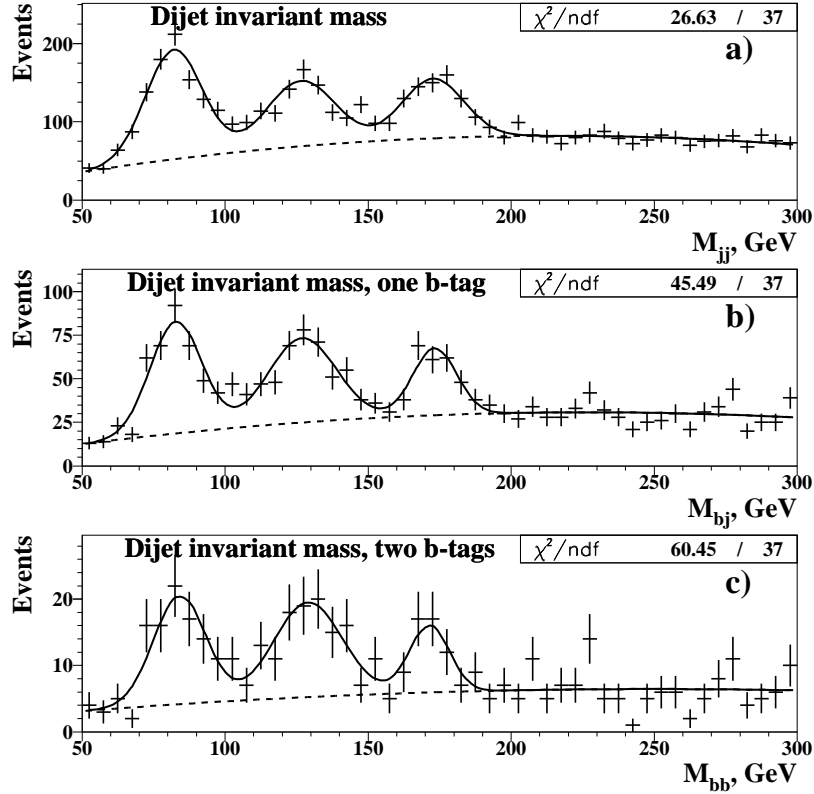}
\caption{Dijet invariant mass observed in black hole decays with a prompt 
lepton or photon tag in $\approx$3~pb$^{-1}$ of the LHC data, for 
$\MP = 1$~TeV and $n=3$: (a) all jet combinations; (b) jet combinations 
with at least one of the jets tagged as a $b$-jet; (c) jet 
combinations with both jets tagged as $b$-jets. The solid line is a 
fit to a sum of three Gaussians and a polynomial background (shown 
with the dashed line). The three peaks correspond to the $W/Z$ bosons, 
the Higgs boson, and the top quark (see text). From \protect\cite{glBH}.}
\label{bhhiggs}
\end{flushright}
\end{figure}

The simplest way to look for the Higgs boson in the black hole decays is to use the 
dijet invariant mass spectrum for all possible combinations of jets found among the final state products in the above sample. This spectrum is shown in figure~\ref{bhhiggs} for $\MP = 1$~TeV and $n=3$. The three panes correspond to all jet combinations (with the average of approximately four jet combinations per event), combinations with at least one $b$-tagged jet, and combinations with both jets $b$-tagged. (We used typical tagging efficiency and mistag probabilities of an LHC detector to simulate $b$-tags.)

The most prominent feature in all three plots is the presence of three peaks with the masses around 85, 130, and 175 GeV. The first peak is due to the hadronic decays of the $W$ and $Z$ bosons produced in the black hole decay either directly or in top-quark decays. (The resolution of a typical LHC detector does not allow to resolve the $W$ and $Z$ in the dijet decay mode.) The second peak is due to the $h \to jj$ decays, and the third peak is due to the $t \to  Wb \to jjb$ decays, where the top quark is highly boosted. In this case, one of the jets from the $W$ decay sometimes overlaps with the prompt $b$-jet from the top quark decay, and thus the two are reconstructed as a single jet; when combined with the second jet from the $W$ decay, this gives a dijet invariant mass peak at the top quark mass. The data set
shown in figure~\ref{bhhiggs} corresponds to 50K black hole events, which, given the 15~nb production cross section for $\MP = 1$~TeV and $n = 3$, is equivalent to the integrated luminosity of 3 pb$^{-1}$, or less than an hour of the LHC operation at the nominal instantaneous luminosity. The significance of the Higgs signal shown in figure~\ref{bhhiggs}a is 6.7$\sigma$, even without $b$-tagging involved.

With this method, a $5\sigma$ discovery of the 130 GeV Higgs boson may be possible with ${\cal L} \approx 2$~pb$^{-1}$ (first day), 100~pb$^{-1}$ (first week), 1~fb$^{-1}$ (first month), 10~fb$^{-1}$ (first year), and 100~fb$^{-1}$ (one year at the nominal luminosity) for the fundamental Planck scale of 1, 2, 3, 4, and 5~TeV, respectively, even with an incomplete and poorly calibrated detector. If the Planck scale is below $\lesssim 4$~TeV, the integrated luminosity required is significantly lower than that for the Higgs discovery in direct production.

While this study was done for a particular value of the Higgs boson mass, the dependence of the new approach on the Higgs mass is small. Moreover, this approach is applicable to searches for other new particles with the masses $\sim 100$~GeV, e.g. low-scale supersymmetry~\cite{SUSYBH}. Light slepton or top squark searches via this technique may be particularly fruitful. Very similar conclusions apply not only to black holes, but to intermediate quantum states, such as string balls~\cite{de}, which have similar production cross section and decay modes as black holes. In this case, the relevant mass scale is not the Planck scale, but the string scale, which determines the Hagedorn evaporation temperature.

Large sample of black holes accessible at the LHC can be used even to study some of the properties of known particles, see, e.g.~\cite{uehara}.

\subsection{Black holes in cosmic rays}

\begin{figure}[thbp]
\begin{flushright}
\epsfxsize=4.2in
\epsffile{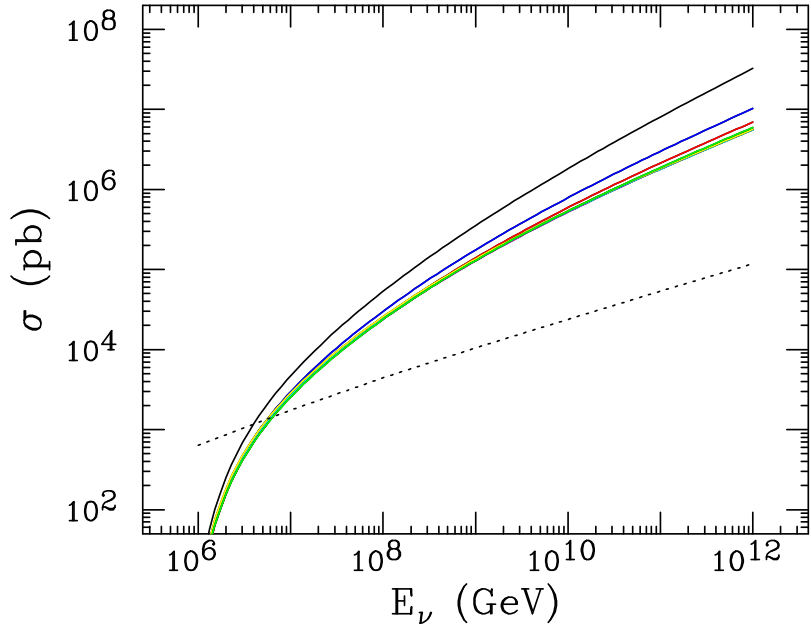}
\caption{Cross sections $\sigma ( \nu N \to {\rm BH})$ for $\MP =
M_{\rm BH}^{\rm min} = 1$~TeV and $n=1, \ldots , 7$. (The last four
curves are virtually indistinguishable.)  The dotted curve is for the
standard model process $\nu N \to \ell X$. From \protect\cite{Feng}.}
\label{bhcr}
\end{flushright}
\end{figure}

Soon after the original papers~\cite{dl,gt} on production of black holes at accelerators have appeared, it has been suggested that similar black hole production can be also observed in the interactions of ultra-high-energy neutrinos with the Earth or its atmosphere~\cite{Feng}. For neutrino energies $\gtrsim 10^7$ GeV, the black hole production cross section in $\nu q$ collisions would exceed their SM interaction rate (see figure~\ref{bhcr}). Several ways of detecting black hole production in neutrino interactions have been proposed~\cite{Feng,CR,afgs,IceCube}, including large-scale ground-based arrays, space probes, and neutrino telescopes as detecting media. 

If the fundamental Planck scale is sufficiently low (1--3 TeV), up to a hundred of black hole events could be observed by, e.g. Pierre Auger observatory even before the LHC turns on. These estimates are based on the so-called guaranteed, or cosmogenic neutrino flux, see, e.g.~\cite{flux}. In certain cosmological models, this flux could be significantly enhanced by additional sources of neutrino emission, e.g. active galactic nuclei; in this case even larger event count is possible.

\begin{figure}[thbp]
\begin{flushright}
\epsfxsize=4.2in
\epsffile{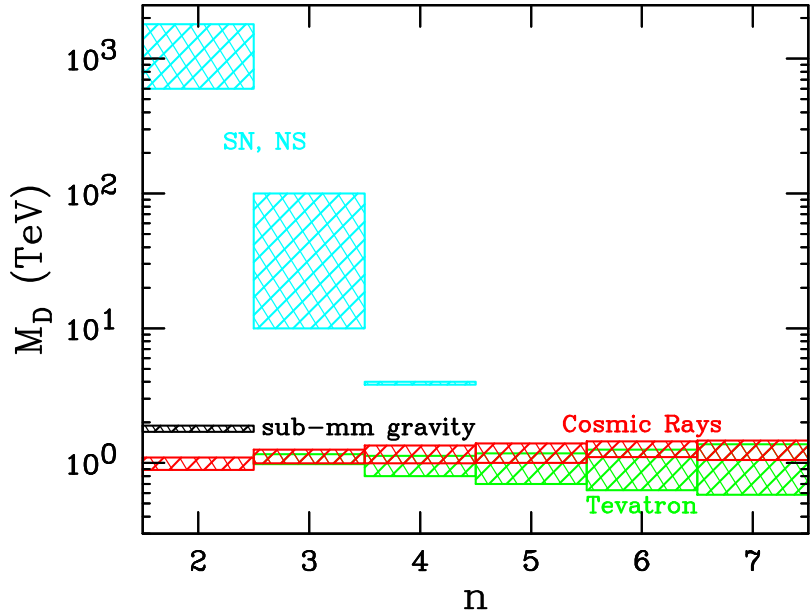}
\caption{Bounds on the fundamental Planck scale from tests of Newton's law on sub-millimeter scales, bounds on supernova cooling and neutron star heating, dielectron and diphoton production at the Tevatron, and non-observation of black-hole production by cosmic neutrinos. For details and estimates of the uncertainties, see \protect\cite{CRlimits}.}
\label{fig:CRlimits}
\end{flushright}
\end{figure}

There are two ways to tell the neutrino interaction that results in a black hole formation from the standard model processes. The first is based on a particular particle content in the black hole events, and would require good particle identification, perhaps beyond the capabilities of the existing detectors. The second approach is based on the comparison of the event rate for Earth-skimming neutrinos (i.e., those that traverse the Earth crust via a short chord, close to the surface) with that for the quasi-horizontal neutrinos (i.e., those that do not penetrate the Earth, but traverse the atmosphere at a small angle). In the former case, many of the neutrinos would be stopped in the Earth due to the large cross section of black hole production. That would suppress the rate of the Earth-skimming-neutrino events in a typical ground array detector, such Pierre Auger. At the same time, the rate of the quasi-horizontal events would increase, as the total cross section, which governs this rate, is dominated by black hole production and therefore is higher than in the standard model case. By measuring the ratio of the two rates, it is possible to distinguish the standard model events from black hole production even with a handful of detected events~\cite{afgs}. As model-dependent as they are, current limits on the fundamental Planck scale derived~\cite{CRlimits} from ultra-high-energy cosmic ray data, are already comparable with those from sub-millimeter gravity measurements~\cite{tabletop} and colliders~\cite{collider}, see figure~\ref{fig:CRlimits}.

Another interesting observation can be done with the IceCube large-scale neutrino telescope at the South Pole, by looking at the zenith angle dependence of the neutrino events at various incidental energies. Similar to the previous argument, significant reduction of the number of observed events due to the neutrino absorption via black hole production in the Earth material surrounding the detector, would occur at smaller zenith angles than that in the case of the standard neutrino interactions~\cite{IceCube}. In addition, particle identification capabilities of the IceCube detector are likely to make it possible to detect the black hole events directly by looking at the event shape. While it has not been mentioned in the original papers~\cite{IceCube}, we would like to note an additional azimuthal dependence of the event rate for high-energy Earth-skimming neutrinos in the IceCube due to the presence of several mountain ridges near the South Pole (particularly, the Transantarctic Mountain Ridge). These mountains are not thick enough to significantly reduce the flux of Earth skimming high-energy neutrinos due to the standard model interactions, but are sufficiently thick to absorb these neutrinos if the black hole creation is allowed. That would constitute a spectacular signature.

\section{Randall-Sundrum black holes}

Mini black holes can be also produced in TeV particle collisions in the Randall-Sundrum model~\cite{rsbh,RizzoGB,RizzoL}. In this case, the warp-factor-suppressed Planck scale, $\Lambda_\pi \sim 1$~TeV, plays the role of the fundamental Planck scale in the model with large extra dimensions.

The event horizon of the Randall-Sundrum black holes has a pancake shape with the radius in the fifth dimension suppressed compared to the radius $R_S$ on the standard model brane by the warp factor $e^{-\pi k R_c}$. Thus, for $R_S e^{-\pi k R_c} \ll \pi R_c$, the black hole can be considered ``small'' and has properties similar to that in the $n = 1$ (5D) large extra dimensions scenario, if the effects of the curvature of the AdS space at the standard model brane are ignored.

In order to derive properties of the Randall-Sundrum black holes, it is convenient to introduce the fundamental 5D Planck scale $M$, which enters the Lagrangian of the Randall-Sundrum model. The relationship between the reduced 4-dimensional Planck scale $\overline{M}_{\rm Pl}$ and $M$ is as follows: $\overline{M}_{\rm Pl}^2 = \frac{M^3}{k}(1 - e^{-2\pi k R_c}) \approx M^3/k$. Since $k$ is 0.01-0.1$\times \overline{M}_{\rm Pl}$, $M = $ 0.2--0.5 $\overline{M}_{\rm Pl}$, i.e. both the 5D and 4D Planck scales are of the same order. Since the curvature of the slice of the AdS space is given by $k^2/M^2 \sim \tilde{k}^2 \ll 1$~\cite{RizzoGB,RizzoL}, one can indeed ignore higher-order curvature effects and consider Randall-Sundrum black holes as if they were black holes in flat Minkowski space.

The Schwarzschild radius of a black hole of mass $M_{\rm BH}$ is given by~\cite{adm,mp,RizzoGB}\footnote{Note that this expression differs from the analogous expression (9) in~\protect\cite{rsbh} by a $\sqrt{2}$ factor; the difference stems from the fact that the mass parameter $M$ used in~\protect\cite{rsbh}is different from the true 5D Planck scale, which enters in the Lagrangian of the model, which we call $M$ in this review.}:
$$
R_S = \frac{1}{\pi M e^{-\pi k R_c}}\sqrt{\frac{M_{\rm BH}}{3M e^{-\pi k R_c}}}.
$$
Taking into account $M^3 \approx k \overline{M}_{\rm Pl}^2 = \Lambda_\pi^2 k e^{2\pi k R_c}$, we get:
\begin{equation}
R_S \approx \frac{1}{\sqrt{3}\pi\Lambda_\pi}\sqrt{\frac{M_{\rm BH}}{\tilde{k}\Lambda_\pi}}.
\label{eq:RSRS}
\end{equation}
Since the expression under the square root is $\sim 10$ for a typical range of $M_{\rm BH}/\Lambda_\pi = O(1)$ and $k/\overline{M}_{\rm Pl} = O(0.01)$, we find that a typical Schwarzschild radius of the Randall-Sundrum black hole is $R_S \sim 1/\Lambda_\pi \sim 1$~TeV$^{-1}$, similar to that for the black holes in models with large extra dimensions. Indeed, using (\ref{eq:RS}) for the ADD model with $n = 1$, we get:
$$
	R_S(\rm ADD, 5D) = \frac{1}{\sqrt{\pi}\MP}\sqrt{\frac{8\MBH}{3\MP}},
$$
which turns into (\ref{eq:RSRS}) for $\Lambda_\pi = \MP$ and $\tilde{k} = 1/8\pi \approx 0.04$.

Moreover, it is easy to see that such a black hole is still small from the point of view of the 5th dimension, as the condition of the black-hole ``smallness'' mentioned above can be expressed as:
$$
	R_S \ll \frac{\pi R_c}{\exp(-\pi k R_c)} = \frac{k\pi R_c}{\frac{k}{\overline{M}_{\rm Pl}}\overline{M}_{\rm Pl}\exp(-\pi k R_c)} = \frac{k\pi R_c}{\Lambda_\pi\tilde{k}} \sim \frac{36}{\Lambda_\pi\tilde{k}}.
$$
Given that $\tilde{k}$ is between 0.01 and 0.1, the inequality becomes:
$$
	R_S \ll \frac{\mbox{360--3600}}{\Lambda_\pi},
$$
which is clearly satisfied for $R_S \sim 1/\Lambda_\pi$. In fact, one would need to produce a black hole with the mass $\sim 10^6$~TeV to exceed this limit. Such energy is achievable neither at any foreseen collider nor in fixed-target collisions of ultra-high-energy particles from cosmic accelerators.

Hawking temperature of the Randall-Sundrum black hole can be found from expression (\ref{eq:TH}) for a black hole in models with large extra dimensions by requiring $n = 1$, i.e. 
\begin{equation}
	T_H = \frac{1}{2\pi R_S},
\label{eq:THRS}
\end{equation}
which, given $R_S \sim 1/\Lambda_\pi$ makes it very similar to that for the case of large extra dimensions. Consequently, for the preferred range of model parameters both the production cross section and the decay properties of a Randall-Sundrum black hole are very similar to those in models with large extra dimensions. In fact, both the TRUENOIR~\cite{truenoir} and CHARYBDIS~\cite{charybdis} generators can be used to simulate Randall-Sundrum black holes by setting $n = 1$ and $\MP = \Lambda_\pi \root 3  \of {8\pi\tilde{k}} \approx 0.765 M_1 \tilde{k}^{-2/3}$.

There has been a suggestion~\cite{rsvsadd} that black holes in Randall-Sundrum model and in large extra dimensions can nevertheless be distinguished by the different dynamics of an early stage of black-hole evaporation due to the fact that the angular momentum of a Randall-Sundrum black hole, unlike that for a black hole in large extra dimensions, cannot have any bulk component (due to the existence of a discrete $Z_2$ orbifold symmetry). Thus, bulk evaporation for a Randall-Sundrum black hole is suppressed compared to that for a black hole of a similar mass in models with large extra dimensions. Since it is argued that the bulk component of Hawking radiation of gravitons for a black hole in large extra dimensions may be significant during the early stages of its evaporation, it is suggested that the black-hole evaporation may result in less missing energy in the Randall-Sundrum scenario. Of course, there are other ways of distinguishing the two cases, particularly the excitation of narrow graviton TeV-scale resonances in Drell-Yan and diboson production in the Randall-Sundrum case~\cite{RSext}, versus an overall enhancement of the high end of the Drell-Yan and diboson mass spectra in the case of large extra dimensions.

Recently, there have been studies of modification of black-hole properties due to Gauss-Bonnet or Lovelock terms added to the Einstein-Hilbert action. These modifications affect properties of black holes in models with either large or warped extra dimension(s). Additional terms could naturally introduce a minimum threshold on the black-hole mass. For detailed studies of modifications related to these higher-order terms, see~\cite{RizzoGB,RizzoL}.

\section{Black holes at CLIC}

A discovery of black holes at the LHC or in ultra-high-energy cosmic ray interactions would open a whole new avenue of studies of quantum gravity. If this is the case, it would be very important to accumulate large samples of black holes in well-controlled conditions, with various initial quantum numbers, including black holes with the quantum numbers of vacuum.

Since most of black holes produced at the LHC would carry electric charge and baryon number, as well as broad range of masses, it is important to have a complementary way of producing completely neutral black holes of fixed mass. An $e^+e^-$ machine of sufficient energy would open such possibility\footnote{A multi-TeV muon collider would offer similar capabilities, although technologically such a machine is more far-fetched.}.

\begin{figure}[tbp]
\begin{flushright}
\epsfxsize=4.2in
\epsffile{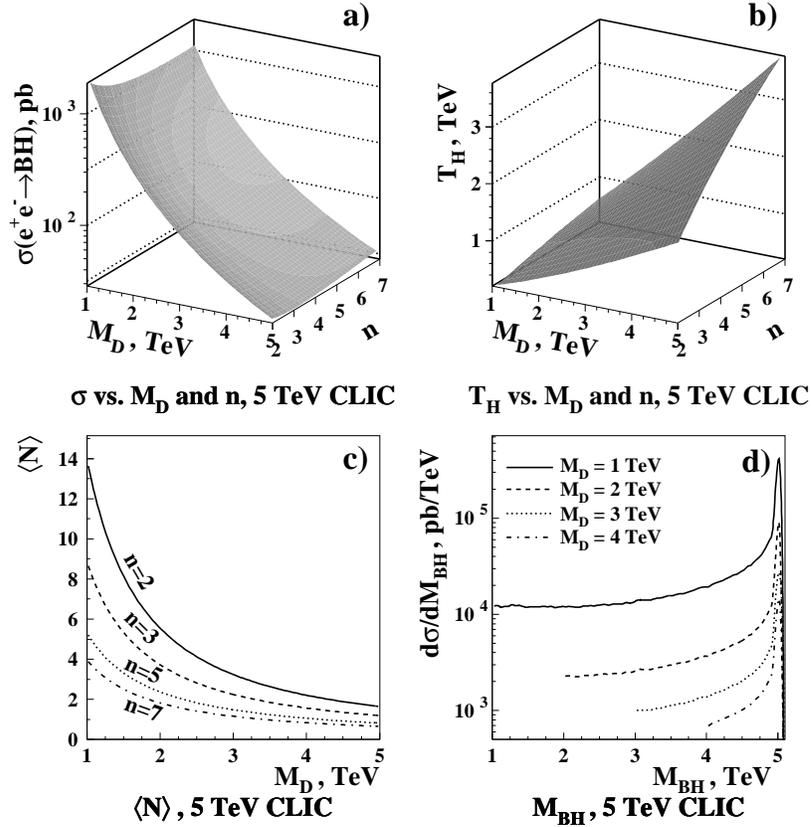}
\caption{Black hole properties at a 5 TeV CLIC $e^+e^-$ collider. Plots a)-c) correspond to the production cross-section, Hawking temperature, and average decay multiplicity for a fixed-mass 5 TeV black hole. Plot d) shows differential cross section of black hole production for $n=4$, as a function of its mass.}
\label{fig:CLIC}
\end{flushright}
\end{figure}

Given the existing limits on the TeV-scale gravity, it is clear that such a machine would have to have multi-TeV energy, i.e. beyond the reach of the proposed International Linear Collider. The only known technology that would allow to build such a machine is two-beam acceleration, which is the basis of a proposed CLIC accelerator with the energy reach of 3--5~TeV~\cite{CLIC}. For completeness, we include properties of black holes produced at a 5~TeV CLIC collider in this review. They are shown in figure~\ref{fig:CLIC}.

There are several advantages of producing black holes in $e^+e^-$ collisions. First of all, unlike that at hadron colliders, nearly all black holes will be produced with the same mass, equal to the total energy of the machine. The spread due to the ``beamstrahlung'' is expected to be small, as shown in figure~\ref{fig:CLIC}d. Second, the black holes are produced at rest and have the vacuum quantum numbers. That would allow to study the details of evaporation process. Third, since the energy of the $e^+e^-$ collider can be readily tuned, one could produce black holes close to the quantum production threshold ($\sim \mp$) and study the effects of quantum gravity near the threshold. Finally, using polarized beams one can change the angular momentum of produced black holes and study how it affects their evaporation. The ability to study all of these fine effects would make CLIC a very attractive complement to the LHC for studies of black holes.

\section{Reentering black holes}

An interesting topic in black hole phenomenology, which has not been studied in much detail yet, is the possibility that a black hole, once produced, moves away into the bulk space. Normally it does not happen as the black holes produced in collisions at the LHC or in cosmic ray interactions are likely to have charge, colour, or lepton/baryon number hair that would keep them on the brane. However, a possibility of that kind is allowed in the case when the strength of gravity in the bulk and on the brane is very different. This is the case, e.g. in the scenario with large extra dimensions with an additional brane term~\cite{braneterm}, or in the case of infinite-volume extra dimensions~\cite{infinite}. 

In these models, a particle produced in a subplanckian collision, e.g. a graviton, could move away in the bulk, where it becomes a black hole due to much lower effective Planck scale in extra dimensions. Since the Planck scale in the bulk is very low, e.g. $\sim 0.01$~eV in the infinite-volume scenario~\cite{infinite}, the newly-formed black hole is very cold and therefore essentially stable. Furthermore, it generally does not move far away from the brane due to gravitational attraction to it, and can further accrete mass from relic energy density in the bulk and from other particles produced in the subsequent collisions. Once the mass of the black hole reaches the mass of the order of the apparent Planck scale, $M_{\rm Pl} \sim 10^{19}$~GeV, the event horizon of the bulk black hole grows so large that it touches the brane, and the black hole immediately evaporates on the brane into $\sim 10$ particles with the energy $\sim 10^{18}$ GeV each. (The energy released in such an event is similar to that in an explosion of a large, few hundred pound conventional bomb!) If such black holes are copiously produced by a remote cosmic accelerator of a reasonable energy, they could act as a source of the highest-energy cosmic rays that are emitted in the process of decay and deceleration of the super-energetic black hole remnants.

Even if the mass of the black hole in the bulk is small, it has certain probability to reenter our brane. In this case, since the event horizon cannot be destroyed, once it has been formed, such a subplanckian object would likely to act as a black hole on the brane and evaporate similarly to a transplanckian black hole discussed above.

\section{Conclusions}

To conclude, black hole production at the LHC and in cosmic rays may be one of the early signatures of TeV-scale quantum gravity in the models with large or warped extra dimensions. It has three advantages:
\begin{enumerate}
\item Large cross section: because no small dimensionless coupling constants, analogous to $\alpha$, suppress the production of black holes; this leads to enormous rates.

\item Hard, prompt, charged leptons and photons: because thermal decays are flavour-blind; this signature has practically vanishing standard model background.

\item Little missing energy: because most of the black hole evaporation products are detectable; this facilitates the determination of the mass and the temperature of the black hole, and may lead to a test of Hawking radiation.
\end{enumerate}

Large samples of black holes accessible by the LHC and the next generation of colliders would allow for precision determination of the parameters of the bulk space and may even result in the discovery of new particles in the black hole evaporation. Limited samples of black hole events may be observed in ultra-high-energy cosmic ray experiments, even before the LHC turns on.

If low-scale gravity is realized in nature, the production and detailed studies of black holes in the lab are just few years away. That would mark an exciting transition for astroparticle physics: its true unification with cosmology~--- the ``Grand Unification'' to strive for.

\ack

I am indebted to my coauthor on the first black hole paper~\cite{dl}, Savas Dimopoulos, for a number of stimulating discussions and support. The idea of reentering black holes is based on numerous communications with Gia Dvali. Many thanks to Tom Rizzo for several enlighting discussions of the properties of Randall-Sundrum black holes. I am grateful to Jonathan Feng for clarifications on the cosmic ray detection of black holes and to Robert Brandenberger and Iain Dell'Antonio for their comments on astrophysical black hole detection. This work has been partially supported by the U.S.~Department of Energy under Grant No. DE-FG02-91ER40688 and by the National Science Foundation under the CAREER Award PHY-0239367. Most of the review was written while the author has been supported by the Fermi National Accelerator Laboratory, as a part of his sabbatical. I wish to thank Fermilab for the hospitality and support.


\section*{References}

\begin{thebibliography}{50}
\bibitem{anthropic}
Susskind L 2005 {\it The Cosmic Landscape: String Theory and the Illusion of Intelligent Design,\/} (Little, Brown)

\bibitem{Riemann}
G.F.B.~Riemann G F B 1868 {\it Abh. Ges. Wiss. G\"ott.\/} {\bf 13} 1

\bibitem{GR}
Einstein A 1907 {\it Jahrbuch der Radioaktivitaet und Elektronik\/} {\bf 4} 411

\nonum
Einstein A 1911 {\it Annalen der Physik\/} {\bf 35} 898

\nonum
Einstein A 1916 {\it Annalen der Physik\/} {\bf 49} 769

\nonum
Einstein A and Grossman M 1913 {\it Z. F. Mathematik und Physik\/} {\bf 62} 225

\bibitem{KK}
Kaluza Th 1921 {\it Sitzungsber. Preuss. Akad. Wiss. Phys. Math. Klasse\/} p~996

\nonum
Klein O 1926 {\it Z.~F.~Physik\/} {\bf 37} 895

\nonum
Klein O 1926 {\it Nature\/} {\bf 118} 516

\bibitem{add}
Arkani-Hamed N, Dimopoulos S and Dvali G 1998 {\it Phys. Lett.\/} B {\bf 429} 263

\nonum
Antoniadis I, Arkani-Hamed N, Dimopoulos S and Dvali G 1998 {\it Phys. Lett.} B {\bf 436} 257

\nonum
Arkani-Hamed N, Dimopoulos S and Dvali G 1999 {\it Phys. Rev.} D {\bf 59} 086004

\bibitem{tabletop}
Hoyle C D {\it et al\/} (E\"ot-Wash Collaboration) 2001 {\it Phys. Rev. Lett.\/} {\bf 86} 1418

\nonum
Adelberger E G (E\"ot-Wash Collaboration) 2002 Sub-mm tests of the gravitational inverse-square law {\it Preprint\/} hep-ex/0202008

\nonum
Chiaverini J, Smullin S J, Geraci A A, Weld D M and Kapitulnik A 2003 {\it Phys. Rev. Lett.\/} {\bf 90} 151101

\nonum
Long J C, Chan H W, Churnside A B, Gulbis E A, Varney M C M and Price J C 2003 
{\it Nature\/} {\bf 421} 922

\nonum
Hoyle C D, Kapner D J, Heckel B R, Adelberger E G, Gundlach J H, Schmidt U and Swanson H E 2004
{\it Phys. Rev.\/} D {\bf 70} 042004

\bibitem{astro}
Cullen S and Perelstein M 1999 {\it Phys. Rev. Lett.\/} {\bf 83} 268 (1999)

\nonum
Hall L and Smith D 1999 {\it Phys. Rev.\/} D {\bf 60} 085008

\nonum
Barger V D, Han T, Kao C and Zhang R J 1999 {\it Phys. Lett.\/} B {\bf 461} 34

\nonum
Fairbairn M 2001 {\it Phys. Lett.\/} B {\bf 508} 335

\nonum
Hanhart C, Pons J A, Phillips D R and Reddy S 2001 {\it Phys. Lett.\/} B {\bf 509} 1

\nonum
Hanhart C, Phillips D R, Reddy S and Savage M J 2001 {\it Nucl. Phys.\/} B {\bf 595} 335

\nonum
Hannestad S and Raffelt G G 2002 {\it Phys. Rev. Lett.\/} {\bf 88} 071301

\nonum
Fairbairn M and Griffiths L M 2002 {\it J. High Energy Phys.\/} JHEP02(2002)024

\bibitem{collider}
Abbott B {\it et al\/} (D\O\ Collaboration) 2001 {\it Phys. Rev. Lett.\/} {\bf 86} 1156

\nonum
Bourilkov D 2001 {\it Proc. 15th Les Rencontres de Physique de la Vallee d'Aoste: Results and Perspective in Particle Physics (La Thuile, Valle d'Aosta, Italy, March 2001)\/} ({\it Prerint\/} hep-ex/0103039)

\nonum
Affolder T {\it et al\/} (CDF Collaboration) 2002 {\it Phys. Rev. Lett.\/} {\bf 89} 281801

\nonum
Abazov V M {\it et al\/} (D\O\ Collaboration) 2003 {\it Phys. Rev. Lett.\/} {\bf 90} 251802

\nonum
Adloff C {\it et al\/} (H1 Collaboration) 2003 {\it Phys. Lett.\/} B {\bf 568} 35

\nonum
Chekanov S {\it et al\/} (ZEUS Collaboration) 2004 {\it Phys. Lett.\/} B {\bf 591} 23

\nonum
Affolder T {\it et al\/} (CDF Collaboration) 2004 {\it Phys. Rev. Lett.\/} {\bf 92} 121802

\nonum
LEP Exotica Working Group, ALEPH, DELPHI, L3, and OPAL Collaborations 2004 {\it CERN Note\/} LEP Exotica WG 2004--03 (URL: {\tt http://lepexotica.web.cern.ch/LEPEXOTICA/notes/2004-03/ed\_note\_final.ps.gz}) and references therein

\nonum
Abazov V M {\it et al\/} (D\O\ Collaboration) 2005 {\it Phys. Rev. Lett.\/} {\bf 95} 161602

\nonum
Abulencia A {\it et al\/} (CDF Collaboration) 2006
{\it Preprint\/} hep-ex/0605101.

\bibitem{hs}
Hewett J L and Spiropulu M 2002 {\it Ann. Rev. Nucl. Part. Sci.\/} {\bf 52} 397

\bibitem{gl}
Landsberg G 2004 {\it Proc. of the 32nd SLAC
Summer Institute on Particle Physics (Menlo Park)} {\it Preprint\/} eConf C040802 MOT006 ({\it Preprint\/} hep-ex/0412028)

\bibitem{RS}
Randall L and Sundrum R 1999 {\it Phys. Rev. Lett.\/} {\bf 83} 3370

\nonum
Randall L and Sundrum R 1999 {\it Phys. Rev. Lett.\/} {\bf 83} 4690

\bibitem{Wise}
Goldberger W D and Wise  M B 1999 {\it Phys. Rev. Lett\/} {\bf 83} 4922

\bibitem{RSext}
Davoudiasl H, Hewett J L and Rizzo T G 2000 {\it Phys. Rev. Lett.\/} {\bf 84} 2080

\nonum
Davoudiasl H, Hewett J L and Rizzo T G 2001 {\it Phys. Rev.\/} D {\bf 63} 075004

\bibitem{RSD0}
Abazov V M {\it et al\/} (D\O\ Collaboration) 2005 {\it Phys. Rev. Lett.\/} {\bf 95} 091801

\bibitem{strangestar}
Drake J J {\it et al.\/} 2002 {\it Astrophys. J.\/} {\bf 572} 996

\bibitem{Ghez} 
Ghez A M {\it et al.\/} 2005 {\it Astrophys. J.\/} {\bf 620} 744

\bibitem{Hawking}
Hawking S W 1975
{\it Commun. Math. Phys.\/} {\bf 43} 199

\bibitem{Oppenheimer}
Tolman R C 1939 {\it Phys. Rev.\/} {\bf 55} 364

\nonum
Oppenheimer J R and Volkov G B 1939 {\it Phys. Rev.\/} {\bf 55} 374
 
\bibitem{cooling}
{\it The Royal Swedish Academy of Sciences Press Release (15 October 1997)\/}
(URL: {\tt http://www.nobelprize.org/physics/laureates/1997/press.html})

\bibitem{Narayan}
Garcia M R, McClintock J E, Narayan R, Callanan P and Murray S S 2000
New evidence for black hole event horizons from Chandra 
{\it Preprint\/} astro-ph/0012452

\nonum
Narayan R and Heyl J S 2002 {\it Astrophys. J.\/} {\bf 574} L139

\bibitem{Narayan1}
Narayan R, Garcia M R and McClintock J E 1997 Advection-dominated accretion and black hole event horizons {\it Preprint\/} astro-ph/9701139

\nonum
Menou K, Quataert E and Narayan R 1997 Astrophysical evidence for black hole event horizons {\it Preprint\/} astro-ph/9712015

\bibitem{adm}
Argyres P C, Dimopoulos S and March-Russell J 1998 {\it Phys. Lett.\/} B {\bf 441} 96

\bibitem{bf}
Banks T and Fischler W 1999
A model for high energy scattering in quantum gravity {\it Preprint\/} hep-th/9906038

\bibitem{ehm}
Emparan R, Horowitz G T and Myers R C 2000
{\it Phys. Rev. Lett.\/} {\bf 85} 499

\bibitem{dl}
Dimopoulos S and Landsberg G 2001 {\it Phys. Rev. Lett.\/} {\bf 87} 161602

\bibitem{gt}
Giddings S B and Thomas S 2002 {\it Phys. Rev.\/} D {\bf 65} 056010

\bibitem{de}
Dimopoulos S and Emparan R 2002 
{\it Phys. Lett.\/} B {\bf 526} 393

\bibitem{mp}
Myers R C and Perry M J 1986 {\it Ann. Phys.\/} {\bf 172} 304

\bibitem{Voloshin}
Voloshin M B 2001 {\it Phys. Lett.\/} B {\bf 518} 137

\nonum
Voloshin M B 2002 {\it Phys. Lett.\/} B {\bf 524} 376

\bibitem{gh}
Gibbons G W and Hawking S W 1977 {\it Phys. Rev.\/} D {\bf 15} 2752

\bibitem{GRcollisions}
Eardley D M and Giddings S B 2002
{\it Phys. Rev.\/} D {\bf 66} 044011

\nonum
Yoshino H and Nambu Y 2002 {\it Phys. Rev.\/} D {\bf 66} 065004

\nonum
Yoshino H and Nambu Y 2003 {\it Phys. Rev.\/} D {\bf 67} 024009

\bibitem{path}
Hsu S D H 2003 {\it Phys. Lett.\/} B {\bf 555} 92

\nonum
Solodukhin S N 2002
{\it Phys. Lett.\/} B {\bf 533} 153

\nonum
Bilke S, Lipartia E and Maul M 2002 Effective field theoretical approach to black hole production {\it Preprint\/} hep-ph/0204040

\bibitem{jt}
Jevicki A and Thaler J 2002
{\it Phys. Rev.\/} D {\bf 66} 024041

\bibitem{EHLQ}
Eichten E, Hinchliffe I, Lane K D and Quigg C 1984 {\it Rev. Mod. Phys.\/} {\bf 56} 579

\bibitem{MRSD}
Martin A D, Roberts R G and Stirling W J 1993 {\it Phys. Lett.\/} B {\bf 306} 145

\nonum
Martin A D, Roberts R G and Stirling W J 1993 {\it Phys. Lett.\/} B {\bf 309} 492 (Erratum)

\bibitem{chromosphere}
Anchordoqui L and Goldberg H 2003 {\it Phys. Rev.\/} D {\bf 67} 064010

\bibitem{grey-body}
Kanti P and March-Russell J 2002
{\it Phys. Rev.\/} D {\bf 66} 024023

\nonum
Kanti P and March-Russell J 2003
{\it Phys. Rev.\/} D {\bf 67} 104019

\nonum
Kanti P, Grain J and Barrau A 2005
{\it Phys. Rev.\/} D {\bf 71} 104002

\nonum
Harris C M 2003
Physics beyond the standard model: exotic leptons and black holes at
future colliders {\it Ph.D Thesis (University of Cambridge)\/} ({\it
Preprint\/} hep-ph/0502005)

\bibitem{grey-body1}
Harris C M and Kanti P 2003 {\it J. High Energy Phys.\/}
JHEP10(2003)014

\bibitem{grey-body2}
Cornell A S, Naylor W and Sasaki M 2006 {\it J. High Energy Phys.\/}
JHEP02(2006)012

\nonum
Cardoso V, Cavaglia M and Gualtieri L 2006
{\it Phys. Rev. Lett.\/} {\bf 96} 071301

\nonum
Cardoso V, Cavaglia M and Gualtieri L 2006 {\it J. High Energy Phys.\/}
JHEP02(2006)021

\nonum
Park D K 2005 Hawking Radiation of the Brane-Localized Graviton from a $(4+n)$-dimensional Black Hole {\it Preprint\/} hep-th/0512021v5

\nonum
Creek S, Efthimiou O, Kanti P and Tamvakis K 2006
{\it Phys. Lett.\/} B {\bf 635} 39

\nonum
Park D K 2006
Emissivities for the various graviton modes in the background of the
higher-dimensional black hole {\it Preprint\/} hep-th/0603224

\bibitem{BH-spin}
Kotwal A V and Hays C 2002
{\it Phys. Rev.\/} D {\bf 66} 091901

\nonum
Ida D, Oda K Y and Park S C 2003
{\it Phys. Rev.\/} D {\bf 67} 064025

\nonum
Ida D, Oda K Y and Park S C 2004 {\it Phys. Rev.\/} D {\bf 69} 049901 (Erratum)

\nonum
Ida D, Oda KY and Park S C 2005
{\it Phys. Rev.\/} D {\bf 71} 124039

\nonum
Park S C and Song H S 2003
{\it J. Korean Phys. Soc.\/} {\bf 43} 300

\nonum
Frolov V P and Stojkovic D 2003
{\it Phys. Rev.\/} D {\bf 67} 084004

\nonum
Frolov V P, Fursaev D V and Stojkovic D 2004 {\it J. High Energy Phys.\/}
JHEP06(2004)057

\nonum
Jung E, Kim S and Park D K 2005
{\it Phys. Lett.\/} B {\bf 615} 273

\nonum
Jung E, Kim S and Park D K 2005
{\it Phys. Lett.\/} B {\bf 619} 347

\nonum
Nomura H, Yoshida S, Tanabe M and Maeda K 2005
{\it Prog. Theor. Phys.\/} {\bf 114} 707

\nonum
Jung E and Park D K 2005
{\it Nucl. Phys.\/} B {\bf 731} 171

\nonum
Duffy G, Harris C M, Kanti P and Winstanley E 2005 {\it J. High Energy Phys.\/}
JHEP09(2005)049

\nonum
Harris C M and Kanti P 2006
{\it Phys. Lett.\/} B {\bf 633} 106

\nonum
Casais M, Kanti P and Winstanley E 2006 {\it J. High Energy Phys.\/}
JHEP02(2006)051

\nonum
Ida D, Oda K Y and Park S C 2006
Rotating black holes at future colliders. III: determination of black hole evolution {\it Preprint\/} hep-th/0602188

\bibitem{microcanonical}
Casadio R and Harms B 2002 {\it Int. J. Mod. Phys.\/} A {\bf 17} 4635

\nonum
Kotwal A V and Hays C 2002 {\it Phys. Rev.\/} D {\bf 66} 116005

\bibitem{recoil}
Frolov V P and Stojkovic D 2002 {\it Phys. Rev.\/} D {\bf 66} 084002

\bibitem{dijets}
Lonnblad L, Sjodahl M and Akesson T 2005 {\it J. High Energy Phys.\/}
JHEP09(2005)019

\nonum
Stocker H 2006 
Stable TeV - black hole remnants at the LHC: discovery through di-jet suppression, mono-jet emission and a supersonic boom in the quark-gluon plasma
{\it Preprint\/} hep-ph/0605062

\bibitem{hadrons}
Mocioiu I, Nara Y and Sarcevic I 2003
{\it Phys. Lett.\/} B {\bf 557} 87

\bibitem{HI}
Chamblin A and Nayak G C 2002
{\it Phys. Rev.\/} D {\bf 66} 091901

\nonum
Chamblin A, Cooper F and Nayak G C 2004
{\it Phys. Rev.\/} D {\bf 69} 065010

\bibitem{inelastic}
Anchordoqui L A, Feng J L, Goldberg H and Shapere A D 2004
{\it Phys. Lett.\/} B {\bf 594} 363

\bibitem{Hagedorn}
Hagedorn R 1965 {\it Nuovo Cimento Suppl.\/} {\bf 3} 147

\bibitem{blackbranes}
Ahn E J, Cavaglia A and Olinto A V 2003 {\it Phys. Lett.\/} B {\bf 551} 1

\nonum
Jain P, Kar S, Panda S and Ralston J 2003 {\it Int. J. Mod. Phys.\/} D {\bf 12} 1593

\nonum
Anchordoqui L A, Feng J L and Goldberg H 2002 {\it Phys. Lett.\/} B {\bf 535} 302

\bibitem{kingman}
Cheung K 2002 {\it Phys. Rev.\/} D {\bf 66} 036007

\bibitem{truenoir}
Dimopoulos S and Landsberg G 2001 {\it Proc. International Workshop on Future of Particle Physics (Snowmass)\/} ({\it Preprint\/} SNOWMASS-2001-P321) (The code and the manual can be found at the following URL: 
{\tt http://hep.brown.edu/users/Greg/TrueNoir/index.htm})

\bibitem{charybdis}
Harris C M, Richardson P and Webber B R 2003 {\it J. High Energy Phys.\/}
JHEP08(2003)033

\bibitem{PYTHIA}
Sj\"{o}strand T {\it et al.\/} 2001 {\it Comput. Phys. Commun.\/} {\bf 135} 238 (We used PYTHIA v6.157)

\bibitem{HERWIG}
G.~Corcella {\it et al.} 2001 {\it J. High Energy Phys.\/}
JHEP01(2001)010

\bibitem{LHC-studies}
Akchurin N, Damgov J, Green D, Kunori S, Landsberg G, Marrafino J, Vidal R, Wenzel H and Wu W 2004
Higgs boson in the Dijet Spectrum of Black Hole Decays at CMS
{\it CMS Internal Note\/} CMS IN--2004/027

\nonum
Harris C M, Palmer M J, Parker M A, Richardson P, Sabetfakhri A and Webber B R 2005 {\it J. High Energy Phys.\/}
JHEP05(2005)053

\nonum
Tanaka J, Yamamura T, Asai S and Kanzaki J 2005
{\it Eur. Phys. J.\/} C {\bf 41} 19

\nonum
Koch B, Bleicher M and Hossenfelder S 2005 {\it J. High Energy Phys.\/}
JHEP10(2005)053

\bibitem{glBH}
Landsberg L 2002 {\it Phys. Rev. Lett.\/} {\bf 88} 181801

\bibitem{SUSYHiggs}
Carena M {\it et al\/} Report of the Higgs working group of the Tevatron run 2 SUSY/Higgs workshop {\it Preprint\/} hep-ph/0010338

\bibitem{SUSYBH}
Chamblin A, Cooper F and Nayak G C 2004
{\it Phys. Rev.\/} D {\bf 70} 075018

\nonum
Nayak G C and Smith J 2006
Higgs boson production from black holes at the LHC {\it Preprint\/} hep-ph/0602129

\bibitem{uehara}
Uehara Y 2002 Black holes at the LHC can determine the spin of Higgs bosons
{\it Preprint\/} hep-ph/0205122

\nonum
Uehara Y 2002 New potential of black holes: quest for TeV-scale physics by measuring top quark sector using black holes {\it Preprint\/} hep-ph/0205199

\bibitem{Feng}
Feng J L and Shapere A D 2002 {\it Phys. Rev. Lett.\/} {\bf 88} 021303

\bibitem{CR}
Anchordoqui L A and Goldberg H 2002
{\it Phys. Rev.\/} D {\bf 65} 047502

\nonum
Emparan R, Masip M and Rattazzi R 2002
{\it Phys. Rev.\/} D {\bf 65} 064023

\nonum
Ringwald A and Tu H 2002 {\it Phys. Lett.\/} B {\bf 525} 135

\nonum
Dutta S I, Reno M H and Sarcevic I 2002
{\it Phys. Rev.\/} D {\bf 66} 033002

\bibitem{afgs}
Anchordoqui L A, Feng J L, Goldberg H and Shapere A D 2002
{\it Phys. Rev.\/} D {\bf 65} 124027

\bibitem{IceCube}
Uehara Y 2002
{\it Prog. Theor. Phys.\/} {\bf 107} 621

\nonum
Kowalski M, Ringwald A and Tu H 2002
{\it Phys. Lett.\/} B {\bf 529} 1

\nonum
Alvarez-Muniz J, Feng J L, Halzen F, Han T and Hooper D 2002
{\it Phys. Rev.\/} D {\bf 65} 124015

\bibitem{flux}
Protheroe R J and Johnson P A 1996 {\it Astropart. Phys.\/} {\bf 4} 253

\nonum
Protheroe R J 1999 {\it Nucl. Phys. Proc. Suppl.\/} {\bf 77} 465

\bibitem{CRlimits}
Anchordoqui L A, Feng J L, Goldberg H and Shapere A D 2003
{\it Phys. Rev.\/} D {\bf 68} 104025

\bibitem{rsbh}
Anchordoqui L A, Goldberg H and Shapere A D 2002
{\it Phys. Rev.\/} D {\bf 66} 024033

\bibitem{RizzoGB}
Rizzo T G 2005 {\it J. High Energy Phys.\/}
JHEP01(2005)028

\nonum
Rizzo T G 2005
TeV-scale black holes in warped higher-curvature gravity
{\it Preprint\/} hep-ph/0510420

\nonum
Rizzo T G 2006
Higher curvature gravity in TeV-scale extra dimensions
{\it Preprint\/} hep-ph/0603242

\bibitem{RizzoL}
Rizzo T G 2005 {\it J. High Energy Phys.\/} JHEP06(2005)079

\nonum
Rizzo T G 2006
TeV-scale black hole lifetimes in extra-dimensional Lovelock gravity
{\it Preprint\/} hep-ph/0601029

\bibitem{rsvsadd}
Stojkovic D 2005 {\it Phys. Rev. Lett.\/} {\bf 94} 011603

\bibitem{CLIC}
Accomando E {\it et al} (CLIC Physics Working Group) 2004 
Physics at the CLIC multi-TeV linear collider {\it Preprint\/} hep-ph/0412251

\bibitem{braneterm}
Dvali G R, Gabadadze G and Kolanovic M 2001 {\it Phys. Rev.\/} D {\bf 64} 084004

\bibitem{infinite}
Dvali G R, Gabadadze G and Kolanovic M 2002 {\it Phys. Rev.\/} D {\bf 65} 024031

\end{thebibliography}
\end{document}